\documentclass[12pt]{article}
\usepackage{amsmath}
\usepackage{graphicx}
\usepackage{enumerate}
\usepackage{natbib}
\usepackage{amsmath}
\usepackage{url} 

\newcommand{\blind}{0}

\DeclareMathAlphabet{\mybm}{OT1}{ptm}{b}{it}

\usepackage{mathptmx}
\usepackage{euscript}
\usepackage{latexsym}
\usepackage{amsmath}
\usepackage{amsfonts}
\usepackage{amssymb}
\usepackage{ulem}
\usepackage{amsthm}

\newtheorem{pro}{Proposition}

{\theoremstyle{remark} 
}

{\theoremstyle{definition} 

}

\newcommand{\BIBO}{{\small \BIBO}}

\newcommand{\plim}{\stackrel{\raisebox{-0.06in}{$\scriptscriptstyle P$\,}}{\rightarrow}}

\newcommand{\linfty}{\rightarrow \infty}
\newcommand{\lzero}{\rightarrow 0}

\newcommand{\eq}{\begin{eqnarray*}}
\newcommand{\eqq}{\end{eqnarray*}}
\newcommand{\eqn}{\begin{eqnarray}}
\newcommand{\eqqn}{\end{eqnarray}}

\newcommand{\eqb}{\begin{align*}}
\newcommand{\eqqb}{\end{align*}}

\newcommand{\tr}{\mathrm{tr}}
\newcommand{\N}{\mathrm{N}}

\newcommand{\eqna}{\begin{align}}
\newcommand{\eqqna}{\end{align}}

\newcommand{\vect}{\mathrm{vec}}

\newcommand{\bA}{\mathbf{A}}
\newcommand{\bB}{\mathbf{B}}
\newcommand{\bD}{\mathbf{D}}

\newcommand{\bH}{\mathbf{H}}
\newcommand{\bI}{\mathbf{I}}

\newcommand{\bQ}{\mathbf{Q}}

\newcommand{\bS}{\mathbf{S}}

\newcommand{\bV}{\mathbf{V}}
\newcommand{\bY}{\mathbf{Y}}
\newcommand{\bX}{\mathbf{X}}
\newcommand{\ba}{\mathbf{a}}

\newcommand{\bx}{\mathbf{x}}
\newcommand{\by}{\mathbf{y}}

\newcommand{\bZ}{\mathbf{Z}}
\newcommand{\bu}{\mathbf{u}}

\newcommand{\bbR}{\mathbb{R}}

\newcommand{\0}{{\bf 0}}

\newcommand{\cF}{\EuScript{F}}

\newcommand{\al}{\alpha}

\newcommand{\ep}{\epsilon}
\newcommand{\gam}{\gamma}

\newcommand{\lam}{\lambda}

\newcommand{\om}{\omega}

\newcommand{\sig}{\sigma}

\newcommand{\bmep}{\pmb{\epsilon}}

\newcommand{\bmbeta}{\pmb{\beta}}
\newcommand{\bmgam}{\pmb{\gamma}}

\newcommand{\bmphi}{\pmb{\phi}}

\newcommand{\bmth}{{\pmb{\theta}}}

\newcommand{\bmxi}{\pmb{\xi}}

\newcommand{\bmzeta}{\pmb{\zeta}}

\newcommand{\bmrho}{\pmb{\rho}}

\newcommand{\bGam}{\pmb{\Gamma}}

\newcommand{\bSig}{\pmb{\Sigma}}

\newcommand{\bPhi}{\pmb{\Phi}}
\newcommand{\bThe}{\pmb{\Theta}}

\newcommand{\bOm}{\pmb{\Omega}}

\newcommand{\albar}{\overline{\al}}
\newcommand{\alu}{\underline{\al}}

\usepackage{float}
\usepackage{caption}
\usepackage{subcaption}
\usepackage{booktabs}

\def\spacingset#1{\renewcommand{\baselinestretch}%
{#1}\small\normalsize} \spacingset{1}

 \DeclareMathSymbol{,}{\mathpunct}{operators}{"2C}

\begin{document}

\if0\blind
{
  \title{\bf  Spline Autoregression Method for Estimation of Quantile Spectrum}
  \author{Ta-Hsin Li\footnote{Formerly affiliated with IBM T.\ J.\ Watson Research Center. 
  Email: {\sc thl024@outlook.com}} }
 \date{August 11, 2025}
  \maketitle
} \fi

\if1\blind
{
  \title{\bf  Spline Autoregression Method for Estimation of Quantile Spectrum} 
  \maketitle
} \fi

\begin{abstract}
Based on trigonometric quantile regression, the quantile spectrum was introduced in Li (2008; 2012) as an alternative tool for spectral analysis of time series. It has been demonstrated to have the capability of providing a richer view of time series data than that offered by the ordinary spectrum especially for nonlinear dynamics such as  stochastic volatility. 
A novel method, called spline autoregression (SAR),  is proposed in this paper for estimating the quantile spectrum as a bivariate function of frequency and quantile level,
under the assumption that the quantile spectrum varies smoothly with the quantile level. 
The SAR method is facilitated by the quantile discrete Fourier transform (QDFT) based on trigonometric quantile regression. It is enabled by the resulting time-domain quantile series (QSER) which represents properly scaled oscillatory characteristics of the original time series around a quantile. A functional autoregressive model is fitted to the QSER on a grid of quantile levels by penalized least-squares, where the autoregressive coefficients are represented as spline functions of the quantile level. While the ordinary autoregressive (AR) model is widely used for conventional spectral estimation, the simulation study in this paper confirms that the proposed SAR method provides an effective way of estimating the quantile spectrum as a bivariate function in comparison with the alternatives.

\vfill
\end{abstract}

\newpage
\spacingset{1.9} 

\section{Introduction}

For a stationary univariate time series $\{ y_t \}$ with continuous
probability distribution function $F(y) := \Pr(y_t \le y)$ and probability 
density function $\dot{F}(y) > 0$, let $r(\tau,\al)$ $(\tau=0,\pm1,\dots)$ denote 
the autocorrelation function of the level-crossing process $\{ I(y_t \le q(\al))\}$, where $I(\cdot)$ is the indicator function and $q(\al) := F^{-1}(\al)$ is the $\al$-quantile of $\{ y_t \}$ for $\al \in (0,1)$.
The quantile spectrum introduced in Li (2008; 2012) through trigonometric quantile regression 
takes the form
\eq
S(\om,\al) := \eta^2(\al) \sum_{\tau=-\infty}^\infty
r(\tau,\al) \exp(-i\om \tau),
\eqq
where $\eta^2(\al) := \al(1-\al) / \dot{F}^2(q(\al))$ is a scaling factor analogous to the variance 
in the conventional spectrum. The quantile spectrum represents the
serial dependence  measured by $r(\tau,\al)$ and retains the information of marginal distribution in $\eta^2(\al)$.
Because $r(\tau,\al)$ is related to the bivariate distribution function $F_\tau(y,y') := \Pr(y_t \le y, y_{t-\tau} \le y')$ such that
\eq
r(\tau,\al) = \frac{1}{\al(1-\al)} \{ F_\tau(q(\al),q(\al))-\al^2\},
\eqq
the quantile spectrum is able to detect nonlinear dependence that cannot be detected by
the conventional spectrum which measures linear dependence  (Li 2012; 2021).

The quantile spectrum can be generalized to the multivariate case (Li 2014a).
Consider $m$ stationary time series $\{y_{j,t} \}$   $(j=1,\dots,m)$ with continuous marginal probability distribution 
functions $F_j(y) := \Pr\{ y_{j,t} \le y\}$ and probability density functions $\dot{F}_j(y) > 0$.
Given a quantile level $\al \in (0,1)$, let $q_j(\al) := F_j^{-1}(\al)$ denote the $\al$-quantile of  $\{ y_{j,t} \}$, 
and  let $r_{jj'}(\tau,\al)$ denote the correlation coefficient between $\{I(y_{j,t} \le q_j(\al))\}$ and $\{I(y_{j',t-\tau} \le q_{j'}(\al))\}$ $(\tau =0,\pm1,\dots)$.
Under the assumption that 
$r_{jj'}(\tau,\al)$ is absolutely summable over $\tau$ for $j,j'=1,\dots,m$ and $\al \in (0,1)$, 
the quantile spectrum $\bS(\om,\al) := [S_{jj'}(\om,\al)]_{j,j'=1}^m$ can be expressed as
\eqn
S_{jj'}(\om,\al) := \eta_j(\al) \eta_{j'}(\al) \sum_{\tau = -\infty} ^\infty r_{jj'}(\tau,\al) \exp(-i \om \tau),
\label{S}
\eqqn
where  $\eta_j(\al) := \sqrt{\al(1-\al)} /\dot{F}_j(q_j(\al))$.
It is easy to show that
\eq
r_{jj'}(\tau,\al)  & = &  \frac{1}{\al(1-\al)} \, \{ F_{jj',\tau}\big(q_j(\al),q_{j'}(\al)\big) - \al^2 \} \\
& = & 1 - \frac{1}{2\al(1-\al)} \, \gam_{jj',\tau}(q_j(\al) , q_{j'}(\al)).
\eqq
In these expressions, $F_{jj',\tau}(y,y') := \Pr\{ y_{j,t} \le y, y_{j',t-\tau} \le y'\}$  denotes  the distribution function
of $(y_{j,t},y_{j',t-\tau})$ and $\gam_{jj',\tau}(y,y') := \Pr\{ (y_{j,t} - y) \, (y_{j',t-\tau}-y' )<0 \}$ denotes the 
 level-crossing rate of $(y_{j,t},y_{j',t-\tau})$. Through these quantities, together with $\eta_j(\al)$, the quantile spectrum in (\ref{S}), as a bivariate function of $\om$ and $\al$, provides a different and richer view 
of  the time series than that offered by the conventional spectrum which is determined
solely by the second moments.

Exploration of the quantile spectrum $\bS(\om,\al)$ as a bivariate function of $\om$  and $\al$
constitutes what we call  quantile-frequency analysis or QFA (Li 2020; 2021; 2023). It contributes to 
a growing literature on nonlinear spectral analysis techniques 
(e.g., Kedem 1986; Hong 2000; Davis and  Mikosch 2009; 
 Hagemann 2013; Dette et al.\ 2015;  Birr et al.\ 2017; 2019;
Fajardo, et al.\  2018; Barun\'{i}k and Kley 2019; Meziani et al.\  2020; Goto et al.\ 2022;
Jordanger and Tj{\o}stheim 2022; 2023; Lim and Oh 2022). 
In this paper, we focus on the estimation of quantile spectrum and refer to earlier publications, e.g., 
Li (2008; 2012; 2014b; 2023; 2025), Chen et al.\ (2022), and Jim\'{e}nez-Var\'{o}n et al.\ (2024), 
for  simulation and real-data examples that illustrate the advantages 
of the quantile spectrum over the conventional spectrum.

Unlike the conventional spectrum of the level-crossing processes (Davis and  Mikosch 2009; Hagemann 2013; Dette et al.\ 2015;  Birr et al.\ 2019;  Goto et al.\ 2022), the quantile spectrum defined by (\ref{S}) retains the information about the marginal distribution through the scaling function $\eta_j(\al)$ as a result of quantile regression.
Furthermore, the quantile spectrum $\bS(\om,\al)$ is treated in this paper as a bivariate function of $\om$ and $\al$ rather than a univariate function of $\om$ for fixed $\al$ as done typically in level-crossing-based 
techniques. 

This paper considers the situation in which $\bS(\om,\al)$ varies smoothly with $\al$. 
For example, $\bS(\om,\al)$ is a continuous function of $\al$ when (a) $\dot{F}_j(F_j^{-1}(\al))$ 
is continuous in $\al$ for all $j$, (b) $F_{jj',\tau}(q_j(\al),q_{j'}(\al))$ is continuous in $\al$ for all $j$, $j'$, 
and $\tau$, and (c) $F_{jj',\tau}(q_j(\al),q_{j'}(\al))-\al^2$ is uniformly summable over $\tau$ for all $j$ and $j'$.
The objective is to leverage 
this smoothness to improve the estimation accuracy over the method that ignores the smoothness 
and estimates the spectrum independent at different quantiles. This line of inquiry was explored
 in Li (2025) using a nonparametric approach. It was demonstrated that leveraging the smoothness 
 across quantiles bear fruits not only for the estimation of quantile spectrum as a bivariate function but also for the estimation of quantile spectrum at a fixed quantile. 

The autoregressive (AR) model is widely used in conventional spectral analysis (Percival and Walden 1993;
Stoica and Moses 1997). This AR approach has been extended by Chen et al.\ (2022) and Jim\'{e}nez-Var\'{o}n et al.\ (2024) to estimate the quantile spectrum. In these works, a two-step procedure is employed 
to produce a bivariate function of $\om$ and $\al$ for estimating the quantile spectrum.
First, an AR model is derived from the quantile periodogram by solving the Yule-Walker equations
for each quantile level in a  set of equally-spaced quantile levels; then the resulting AR parameters are smoothed nonparametrically across the quantile levels.

In this paper, we  propose a new method that combines autoregression and quantile smoothing
into a single procedure. This method, called spline autoregression (SAR), 
 is enabled by what we call the quantile series (QSER). Each QSER is derived from the original time series through what we call the quantile discrete Fourier transform (QDFT) at a quantile level based on trigonometric quantile regression. It is a time series whose ordinary periodogram
 coincides with the quantile periodogram. The QSER replaces the quantile periodogram
 used in Chen et al.\ (2022) and Jim\'{e}nez-Var\'{o}n et al.\ (2024) to derive the AR model in the
 proposed SAR method. More specifically, the SAR method fits a functional AR model to the QSER 
 by penalized least-squares, where the AR parameters are represented as spline functions 
 of $\al$ and the roughness of these functional coefficients are penalized. 
Our simulation study shows that the SAR method is able to produce more accurate estimates than the alternatives which apply autoregression independently to the QSER at each quantile level, even with subsequent smoothing across quantiles. 

The rest of the paper is organized as follows. Section 2 introduces the QDFT and the resulting quantile periodogram (QPER); Section 3 introduces the QSER and the SAR estimator.
Section~4 discusses some properties of the SAR estimator. 
Section 5 presents the result of a simulation study that evaluates the proposed method.
Concluding remarks are given in Section 6. 

Supplementary material to the main body of the paper includes 
discussions on the computation of the SAR estimator (Appendix I)
and its application to Granger-causality analysis (Appendix II), additional simulation results (Appendix III), and
a summary of R functions that implement the proposed method in the R package `qfa' (Appendix IV).

\section{Quantile Fourier Transform and Quantile Periodogram}

Given a data  record $\{ y_{j,t} : t=1,\dots,n \}$ ($j=1,\dots,m$),  consider the quantile regression problem
\eqn
\hat{\bmbeta}_j(\om,\al) := \operatorname*{argmin}_{\bmbeta}
\sum_{t=1}^n \rho_{\al} \big(y_{j,t} - \bx_{t}^T(\om) \, \bmbeta \big), 
\label{qr}
\eqqn
where $\om \in [0,2\pi)$ is the frequency variable, $\al \in (0,1)$ is the quantile level variable, 
$\rho_\al(y) := y (\al - I(y \le 0))$ is the objective function of quantile regression 
(Koenker 2005, p.\ 5), and $\bx_t(\om)$ is a  trigonometric regressor defined by
\eqn
\bx_t(\om)  := 
\left\{ 
\begin{array}{ll}
1 &  \om=0, \\
\text{$[1,\cos(\pi t)] ^T$} & \om = \pi, \\
\text{$[1,\cos(\om t),\sin(\om t)]^T$} &  \text{otherwise}.
\end{array} \right.
\label{xt}
\eqqn
The solution of (\ref{qr}) takes the form
\eqn
\hat{\bmbeta}_j(\om,\al)  = 
\left\{ 
\begin{array}{ll}
 \hat{\beta}_{1,j}(0,\al) &  \om =0, \\
\text{$[\hat{\beta}_{1,j}(\pi,\al),\hat{\beta}_{2,j}(\pi,\al)] ^T$} & \om = \pi, \\
\text{$[\hat{\beta}_{1,j}(\om,\al),\hat{\beta}_{2,j}(\om,\al),\hat{\beta}_{3,j}(\om,\al)]^T$} &  
\text{otherwise}.
\end{array} \right.
\label{beta}
\eqqn
Based on these quantities, we define
\eqn
Z_j(\om,\al)  := 
\left\{ 
\begin{array}{ll}
n \, \hat{\beta}_{1,j}(0,\al) & \om=0, \\
n \, \hat{\beta}_{2,j}(\pi,\al) & \om = \pi, \\
(n/2) \{ \hat{\beta}_{2,j}(\om,\al) - i \, \hat{\beta}_{3,j}(\om,\al) \} & \mbox{otherwise},
\end{array} \right.
\label{qdft}
\eqqn
where $i := \sqrt{-1}$. The quantile periodogram (QPER) introduced in Li (2028; 2012; 2014a) can be expressed as
$\bQ(\om,\al) := [Q_{jj'}(\om,\al)]_{j,j'=1}^m$, where
\eqn
Q_{jj'}(\om, \al) := n^{-1} Z_j(\om,\al) \, Z_{j'}^*(\om,\al) \quad (j,j'=1,\dots.m),
\label{qper}
\eqqn
with superscript $*$ denoting complex conjugate.
Under suitable conditions (Li 2014a, p.~557), it can be shown that
for any fixed $0 < \lam_1 < \dots < \lam_q < \pi$, 
$\bQ(\lam_1,\al),\dots,\bQ(\lam_q,\al)$ are asymptotically distributed as 
$\bmzeta_1 \bmzeta_1^H,\dots,\bmzeta_q \bmzeta_q^H$ (with superscript $H$ denoting Hermitian transpose), 
where $\bmzeta_1,\dots,\bmzeta_q$ are independent complex Gaussian random vectors
with zero mean and covariance matrix $\bS(\lam_1,\al),\dots,\bS(\lam_q,\al)$, respectively.
This is analogous to a property of the conventional periodogram in which $\bS(\om,\al)$ plays the role of 
the conventional spectrum (Brockwell and Davis 1991, p.~446).

Let $\om_v := 2\pi v/n$ $(v=0,1,\dots,n-1)$ be the $n$ Fourier frequencies. 
For each $j \in \{1,\dots,m\}$, the sequence $\{ Z_j(\om_v,\al): v=0,1,\dots,n-1\}$ 
constitutes what we call the quantile discrete Fourier transform (QDFT) of $\{ y_{j,t}: t=1,\dots,n \}$ 
at quantile level $\al$  (Li 2025). The QDFT is an extension of the conventional DFT, $Z_j(\om_v) : = \sum_{t=1}^n y_{j,t} \exp(-i \om_v t)$ $(v=0,\dots,n-1)$, because the latter can be obtained from the trigonometric regression machinery  (\ref{qr})--(\ref{qper}) with $\rho_\al(y)$ replaced by $y^2$. Similarly, 
the quantile periodogram $\{\bQ(\om_v,\al): v=0,1,\dots,n-1\}$ is  an extension 
of the conventional periodogram $\{  [  I_{jj'}(\om_v) ]_{j,j'=1}^m: v=0,1,\dots,n-1\}$, where
$I_{jj'}(\om_v) := n^{-1} Z_j(\om_v) \,  Z_{j'}^*(\om_v)$. Unlike some fast Fourier transform algorithms 
that compute the conventional DFT, the trigonometric quantile regression method 
that computes the QDFT does not require $n$ to be a positive integer  power of 2 or the zero-padding technique 
to make it so.

\section{Quantile Series and Spline Autoregression Estimator}

Associated with the QDFT is a time-domain series, which we call
the quantile series (QSER), defined as the inverse discrete Fourier transform of the QDFT (Li 2025), i.e.,
\eqn
y_{j,t}(\al) := n^{-1} \ \sum_{v=0}^{n-1} Z_j(\om_v,\al) \exp(i t \om_v) \quad (t=1,\dots,n).
\label{qser}
\eqqn
This is a  real-valued sequence with mean $\bar{y}_j(\al) := n^{-1} \sum_{t=1}^n y_{j,t}(\al)$ equaling  
 the $\al$-quantile of $\{ y_{j,t}: t=1,\dots,n\}$, i.e., $\bar{y}_j(\al) = \hat{\beta}_j(0,\al)$.
The quantile periodogram $\{\bQ(\om_v,\al): v=0,1,\dots,n-1\}$ coincides 
with the conventional periodogram of $\{ y_{j,t}(\al): t=1,\dots,n\}$. This observation gives rise to the idea of applying
conventional spectral estimation techniques to the QSER in developing different estimators 
for the quantile spectrum.

As discussed in Li (2025), the QSER can be viewed as a surrogate of the underlying 
quantile-crossing process
\eqn
u_{j,t}(\al) := q_j(\al) +  \{ \al - I( y_{j,t} \le q_j(\al)) \} /\dot{F}_j(q_j(\al)).
\label{qc}
\eqqn
While a rigorous proof remains 
open for future research, a heuristic argument can be made to justify the QSER by using the Bahadur-type representations 
(Wu 2007; Li 2012)
\eq
 \hat{\beta}_{1,j}(0,\al) & = & n^{-1}  \sum_{t=1}^n  u_{j,t}(\al) + o_P(n^{-1/2}), \\
  \hat{\beta}_{2,j}(\pi,\al) & = &  n^{-1}  \sum_{t=1}^n  u_{j,t}(\al) \cos(\pi t) + o_P(n^{-1/2}), \\
 \hat{\beta}_{2,j}(\om_{v},\al) & = & 2   \, n^{-1} \sum_{t=1}^n u_{j,t}(\al) \cos(\om_{v} t) + 
o_P(n^{-1/2}) \quad 
 \om_{v} \notin \{ 0,\pi\}, \\
 \hat{\beta}_{3,j}(\om_{v},\al) & = & 2 n^{-1} \sum_{t=1}^n u_{j,t}(\al) \sin(\om_{v} t) 
+ o_P(n^{-1/2}) \quad \om_{v} \notin \{0,\pi\}.
\eqq
If these expressions hold uniformly, then it follows  that
\eq
y_{j,t}(\al) = u_{j,t}(\al) + e_{j,t}(\al)
\eqq
and $n^{-1} \sum_{t=1}^n \{e_{j,t}(\al) \}^2= o_P(1)$. In other words, the QSER should behave like 
an approximation to the underlying quantile-crossing process.
The latter is stationary with a conventional spectrum that coincides with the quantile spectrum $\bS(\om,\al)$.

Let $\bGam(\tau,\al)$ denote the autocovariance function (ACF) of 
$\bu_t(\al) := [u_{1,t}(\al),\dots,u_{m,t}(\al)]^T$ . Then,
the quantile spectrum $\bS(\om,\al)$ can be expressed as
\eq
\bS(\om,\al) = \sum_{\tau=-\infty}^\infty \bGam(\tau,\al) \, \exp(- i \om \tau).
\eqq
This relationship was exploited in Li (2025) to device the lag-window (LW) estimator
\eqn
\hat{\bS}_{\rm LW}(\om,\al)   =
\sum_{|\tau| \le M}  h(\tau/M) \, \hat{\bGam}(\tau,\al) \exp(- i \om \tau ),
\label{lw}
\eqqn
where $h(\cdot)$ is a window function  and $\hat{\bGam}(\tau,\al)$ is the ACF of $\{ y_{j,t}(\al): t=1,\dots,n \}$ $(j=1,\dots,m)$,
called the quantile ACF (QACF), which serves as an estimate of $\bGam(\tau,\al)$. 

In the following, we focus on  the AR approach. Let $\bA_p(\al) := [\bA_{p,1}(\al),\dots,\bA_{p,p}(\al)]$ 
denote the matrix of AR coefficients obtained from the Yule-Walker equations of an AR$(p)$ process whose ACF is 
$\bGam(\tau,\al)$ $(\tau=0,\pm 1,\dots)$, and let $\bV_p(\al) := [\sig_{jj'}(\al) ]_{j,j'=1}^m$ denote the resulting residual covariance matrix. It can be shown (L\"{u}tkepohl 1993, p.\ 78) that
\eqn
\bA_p(\al) = \bmgam_p(\al) \, \bGam_p^{-1}(\al), \quad 
\bV_p(\al) =  \bGam(0,\al) - \bA_p(\al) \, \bmgam_p^T(\al),
\label{yw}
\eqqn
where $\bGam_p(\al) := [\bGam(\tau-\tau',\al)]_{\tau,\tau'=1}^p$ and $\bmgam_p(\al) := [ \bGam(1,\al),\dots,\bGam(p,\al)]$. Associated with these parameters is the AR spectrum 
\eqn
\bS_p(\om,\al) := (\bI - \bA_p(\om,\al))^{-1} \bV_p(\al)  \, (\bI - \bA_p(\om,\al))^{-H},
\label{AR}
\eqqn
with $\bA_p(\om,\al)  := \sum_{\tau=1}^p \bA_{p,\tau} \exp(- i \om \tau)$. (Superscript $-H$ stands for the Hermitian transpose of the inverse of a matrix.) This spectrum has the maximum entropy property (Parzen 1982; Choi 1993), i.e., it maximizes the entropy among all spectra whose first $p+1$ autocovariances
coincide with $\{ \bGam(\tau,\al): \tau=0,1,\dots,p\}$. We employ this spectrum as a model to approximate 
$\bS(\om,\al)$. The error of approximation  satisfies
\eqn
\| \bS_p(\om,\al) - \bS(\om,\al) \|_1
\le \sum_{|\tau| > p} \| \bGam_p(\tau,\al) \|_1 + \sum_{|\tau| > p} \| \bGam(\tau,\al) \|_1,
\label{err}
\eqqn
where $\bGam_p(\tau,\al)$ is the ACF of the AR model satisfying $\bS_p(\om,\al) 
= \sum_{\tau=-\infty}^\infty \bGam_p(\tau,\al) \exp(-i \om \tau)$.
The second term on the right-hand side of (\ref{err}) tends to zero as $p \linfty$, due to the 
assumption that all entries in $\bGam(\tau,\al)$ are absolutely summable over $\tau$.
If  the first term also tends to zero, we would have $\| \bS_p(\om,\al) - \bS(\om,\al) \|_1 
\lzero$ as $p \linfty$, which justifies the AR approach. 
Another way of justifying this  approach is through the AR($\infty$) representation of $\bS(\om,\al)$ as the ordinary spectrum of $\{ \bu_t(\al) \}$ (Wiener and Masani 1957; 1958; Whittle 1963).

 For fixed $\al$, the AR parameters in (\ref{AR}) can be estimated by solving the Yule-Walker
equations with $\{ \bGam(\tau,\al): \tau=0,1,\dots,p\}$ replaced 
by $\{ \hat{\bGam}(\tau,\al): \tau=0,1,\dots,p\}$. To obtain an estimate over a given interval of
$\al \in [\alu,\albar] \subset (0,1)$, one may first compute the AR parameters 
for each quantile level in a set of equally-spaced quantile levels $\al_1 := \alu < \al_2 < \dots < \al_L := \albar$,
and then apply a smoothing procedure to the resulting AR parameters across the 
quantile levels $\{ \al_\ell: \ell=1,\dots,L\}$.
This two-step method has been explored by Chen et al.\ (2022) and Jim\'{e}nez-Var\'{o}n et al.\ (2024) with
$\hat{\bGam}(\tau,\al)$ derived from the quantile periodogram instead of the QSER.

We propose a new method for estimating the AR parameters in (\ref{AR}) as functions of $\al \in [\alu,\albar]$. This method, called spline autoregression (SAR),  is based on least-squares
autoregression which is applied jointly to the QSER for all $\al_\ell$ $(\ell=1,\dots,L)$.
The AR parameters in (\ref{AR}) are represented as spline functions of $\al \in [\alu,\albar]$
and penalized for their roughness in the least-squares procedure.

More precisely, let $\by_t(\al_\ell) := [y_{1,t}(\al_\ell)-\bar{y}_1(\al_\ell),\dots,y_{m,t}(\al_\ell)-\bar{y}_m(\al_\ell)]^T$ 
$(t=1,\dots,n)$ denote the demeaned QSER at $\al_\ell$ $(\ell=1,\dots,L)$. Then, the SAR problem can be stated as
\eqn
\{ \hat{\bA}_1(\cdot),\dots,\hat{\bA}_p(\cdot)\}  & := & 
\operatorname*{argmin}_{\bA_1(\cdot),\dots,\bA_p(\cdot) \in \cF_m} \
\sum_{\ell=1}^L (n-p)^{-1} \sum_{t=p+1}^n
\bigg\| \by_t(\al_\ell) - \sum_{\tau=1}^p \bA_\tau(\al_\ell) \, \by_{t-\tau}(\al_\ell) \bigg\|^2 \notag \\
& & \qquad \qquad \qquad +\  \lam  \sum_{\tau=1}^p  \int_{\alu}^{\albar} \| \ddot{\bA}_\tau(\al) \|^2 \, d\al,
\label{sar}
\eqqn
where $\cF_m$ is the space spanned by $m$-by-$m$ 
matrices of spline basis functions in $[\alu,\albar]$ and
$\lam > 0$ is the smoothing parameter that specifies the amount of penalty for the roughness 
of the functional AR coefficients as measured by the integral of the Frobenius norm of second derivatives.  In addition, let the residual covariance matrix at $\al_\ell$  be defined by
\eq
\tilde{\bV}(\al_\ell)  & := & [\tilde{\sig}_{jj'}(\al_\ell)]_{j,j'=1}^m  \\
& := & (n-p)^{-1} \sum_{t=p+1}^n  \bigg[\by_t(\al_\ell) - \sum_{\tau=1}^p \hat{\bA}_\tau(\al_\ell) \, \by_{t-\tau}(\al_\ell) \bigg] \notag \\
& &  \hspace{1in} 
\times \bigg[\by_t(\al_\ell) - \sum_{\tau=1}^p \hat{\bA}_\tau(\al_\ell) \, \by_{t-\tau}(\al_\ell) \bigg]^T.
\eqq
Applying the smoothing spline technique to $\{ \tilde{\sig}_{jj'}(\al_\ell) : \ell=1,\dots,L \}$ yields
\eqn
\hat{\bV}(\cdot) := [\hat{\sig}_{jj'}(\cdot)]_{j,j'=1}^m.
\label{Vhat}
\eqqn
where
\eqn
\hat{\sig}_{jj'}(\cdot) := 
\operatorname*{argmin}_{\sig(\cdot) \in \cF_1} \
\bigg\{
\sum_{\ell=1}^L (\tilde{\sig}_{jj'}(\al_\ell) - \sig(\al_\ell))^2
+ \lam \int_{\alu}^{\albar} (\ddot{\sig}(\al))^2 d\al \bigg\}.
\label{sighat}
\eqqn
Note that we set the smoothing parameter in (\ref{sighat}) to be identical to 
the smoothing parameter in  (\ref{sar}) for consistency and simplicity, although it could take a different value $\lam_{jj'}$ in general.

The SAR solution given by (\ref{sar})--(\ref{sighat}) can be viewed 
as a smoothing spline estimate of  the AR parameters $\{ \bA_{p,1}(\cdot),\dots,
\bA_{p,p}(\cdot)\}$ and $\bV_p(\cdot)$  in (\ref{AR}) defined by the Yule-Walker equations.
Plugging these estimates in (\ref{AR}) leads to the SAR spectral estimator
\eqn
\hat{\bS}(\om,\al)
:= (\bI - \hat{\bA}(\om,\al))^{-1} \hat{\bV}(\al) \,  (\bI - \hat{\bA}(\om,\al))^{-H},
\label{Shat}
\eqqn
where $\hat{\bA}(\om,\al) := \sum_{\tau=1}^p \hat{\bA}_{\tau}(\al) \exp(-i\tau\om)$ and $(\om,\al) \in [0,2\pi) \times [\alu,\albar]$.

To complete the procedure, we propose a data-driven method for selecting $p$ and $\lam$ in (\ref{sar}): First,  for each $\ell=1,\dots,L$, fit an  AR($p$) model to $\{ \by_t(\al_\ell): t=1,\dots,n \}$ to obtain Akaike's information criterion $\text{AIC}_p(\al_\ell)$ with $p=0,1,\dots,p_0$ (L\"{u}tkepohl 1993, p.\ 129), where $p_0$ is a pre-determined maximum order. Then, choose $p$ to minimize 
the average AIC across the quantile levels, i.e., $L^{-1} \sum_{\ell=1}^L \text{AIC}_p(\al_\ell)$. Finally,
with $p$ given and fixed, choose $\lam$ to minimize the generalized cross-validation (GCV) criterion 
\eqn
\text{GCV}(\lam) := \frac{ (L(n-p))^{-1}
\sum_{\ell=1}^L \sum_{t=p+1}^n
\| \by_t(\al_\ell) - \sum_{\tau=1}^p \hat{\bA}_\tau(\al_\ell) \, \by_{t-\tau}(\al_\ell) \|^2} 
{\{1-(L(n-p))^{-1} \tr({\bH})\}^2},
\label{gcv}
\eqqn
where $\tr(\bH)$ is the trace of the hat  matrix associated with (\ref{sar}) (see Appendix I for details),
which serves as the effective degree of freedom (Hastie and Tibshirani 1990, p.\ 52).  Note that selecting  $\lam$ and $p$ jointly to minimize the GCV in (\ref{gcv}) is problematic, because 
the response of $\tr(\bH)$  to over-parameterization with a large $p$ 
can be mitigated by  the choice of a large $\lam$. The proposed method of selecting $p$ independently before  $\lam$
overcomes this difficulty.

\section{Characterization of the Spline Autoregression Estimator}

Let $\cF_m$ be the space spanned by $m$-by-$m$ matrices of spline basis functions 
$\{ \phi_k(\cdot): k=1,\dots,K \}$. For any $\bA_\tau(\cdot) \in \cF_m$,
there exists $\bThe_\tau :=  [ \bThe_{\tau,1},\dots, \bThe_{\tau,K}] \in \bbR^{m \times Km}$ 
such that
\eq
\bA_\tau(\cdot) := \sum_{k=1}^K  \bThe_{\tau,k} \, \phi_k(\cdot) = \bThe_{\tau} \, \bPhi(\cdot),
\eqq
where $\bPhi(\cdot) :=  [ \phi_1(\cdot)\bI_m,\dots,\phi_K(\cdot)\bI_m]^T \in \bbR^{Km \times m}$. Define
\eq
\bY_\ell & := & [\by_{p+1}(\al_\ell),\dots,\by_n(\al_\ell)]  \in \bbR^{m \times (n-p)}, \\[0.1in]
\bZ_\ell  & := &
\left[
\begin{array}{ccc}
 \bPhi(\al_\ell) \, \by_p(\al_\ell) & \cdots &  \bPhi(\al_\ell) \, \by_{n-1}(\al_\ell) \\
\vdots &         & \vdots \\
 \bPhi(\al_\ell) \, \by_1(\al_\ell) & \cdots &  \bPhi(\al_\ell) \, \by_{n-p}(\al_\ell)
\end{array}
\right] \in \bbR^{Kmp \times (n-p)}, \\
\bD  & := & \bI_p \otimes  \int_{\alu}^{\albar}  \ddot{\bPhi}(\al) \,\ddot{\bPhi}^T(\al) \, d\al  
\in  \bbR^{Kmp \times Kmp}.
\eqq
With this notation, we have the following results for the SAR solution in (\ref{sar}).

\begin{pro}
{\rm
\begin{itemize}
\item[(a)]
The solution of (\ref{sar}) can be expressed as 
\eq
\hat{\bA}(\cdot)  := [\hat{\bA}_1(\cdot),\dots,\hat{\bA}_p(\cdot)]= [ \hat{\bThe}_1 \, \bPhi(\cdot),\dots,\hat{\bThe}_p \, \bPhi(\cdot)] = 
\hat{\bThe} \, (\bI_p \otimes \bPhi(\cdot)),
\eqq
where
\eqn
\hat{\bThe}  := [ \hat{\bThe}_{1},\dots, \hat{\bThe}_{p}] 
:= \bigg( \sum_{\ell=1}^L \bY_{\ell} \bZ_{\ell}^T \bigg)
\bigg( \sum_{\ell=1}^L \bZ_\ell \bZ_\ell^T  + (n-p) \lam \bD  \bigg)^{-1}.
\label{Theta}
\eqqn
\item[(b)]
As $n \linfty$, if $\hat{\bGam}(\tau,\al) \plim \bGam(\tau,\al)$ for any fixed $\tau$ and $\al$, then 
 $\hat{\bA}(\al) \plim \bar{\bA}(\al)  := \bar{\bThe} \, (\bI_p \otimes \bPhi(\al))$ 
 uniformly in $\al \in [\alu,\albar]$, where 
\eq
\bar{\bThe}  & := & \bigg( \sum_{\ell=1}^L
\bA_p(\al_\ell) \, \bGam_p(\al_\ell)  \, (\bI_p \otimes \bPhi^T(\al_\ell))
\bigg)   \\
& & \times 
\bigg(
\sum_{\ell=1}^L  (\bI_p \otimes \bPhi(\al_\ell)) \, \bGam_p(\al_\ell) \, (\bI_p \otimes \bPhi^T(\al_\ell)
+ \lam \, \bD \bigg) ^{\!\!\! -1}.
\eqq
\item[(c)]
If\/ $\bA_{p,\tau}(\cdot) \in \cF_m$ for all $\tau=1,\dots,p$, then, as $\lam \lzero$, $\bar{\bA}(\al) \rightarrow \bA_p(\al)$ uniformly in $\al \in [\alu,\albar]$.
\end{itemize}
}
\label{thm:a}
\end{pro}

\noindent
{\sc Proof}: The proof of assertion (a) can be found in Appendix I. 
To prove assertion (b), we note that $\hat{\bGam}(\tau,\al) \plim \bGam(\tau,\al)$ implies
\eq
(n-p)^{-1} \bZ_\ell \bZ_\ell^T & \plim &
 [ \bPhi(\al_\ell) \, \bGam(\tau-\tau',\al_\ell) \, \bPhi^T(\al_\ell)]_{\tau,\tau'=1}^p \\
& = & (\bI_p \otimes \bPhi(\al_\ell)) \, \bGam_p(\al_\ell) \, (\bI_p \otimes \bPhi^T(\al_\ell)), \\
(n-p)^{-1} \bY_\ell \bZ_\ell^T & \plim & [ \bGam(1,\al_\ell) \, \bPhi^T(\al_\ell),\dots,\bGam(p,\al_\ell) \, \bPhi^T(\al_\ell)] \\
& = & \bmgam_p(\al_\ell) \, (\bI_p \otimes \bPhi^T(\al_\ell)).
\eqq
Combining this  with (\ref{yw})  and (\ref{Theta}) proves $\hat{\bThe} \plim \bar{\bThe}$. Hence the assertion (b). 
Under the condition in (c), there exists
$\bThe  \in \bbR^{m \times Kmp}$ such that $\bA_p(\cdot) = \bThe (\bI_p \otimes\bPhi(\cdot))$.
In this case, 
\eq
\bar{\bThe}  & = & \bThe \, \bigg(  \sum_{\ell=1}^L
(\bI_p  \otimes \bPhi(\al_\ell) ) \, \bGam_p(\al_\ell) \, (\bI_p \otimes \bPhi^T(\al_\ell))
\bigg) \\
& & \times  \bigg(
\sum_{\ell=1}^L  (\bI_p \otimes \bPhi(\al_\ell)) \, \bGam_p(\al_\ell) \, (\bI_p \otimes \bPhi^T(\al_\ell)
+ \lam \, \bD \bigg)^{\!\!\!-1}.
\eqq
As $\lam \lzero$, we have $\bar{\bThe} \rightarrow \bThe$. Therefore, $\bar{\bA}(\al) 
\rightarrow \bThe\, (\bI_p \otimes\bPhi(\al))$ uniformly in $\al \in [\alu,\albar]$. \qed

 \medskip
Similarly, the following results can be obtained for $\hat{\bV}(\cdot)$ in (\ref{Vhat}) with the notation $\bmphi(\al) := [\phi_1(\al),\dots,\phi_K(\al)]^T$, $\bB := [\bmphi(\al_1),\dots,\bmphi(\al_L)]^T$, and $\bOm := 
\int_{\alu}^{\albar} \ddot{\bmphi}(\al) \,  \ddot{\bmphi}^T(\al) \, d\al$. 
 
\begin{pro} 
{\rm
\begin{itemize}
\item[(a)] The solution of (\ref{sighat}) can be expressed 
as $\hat{\sig}_{jj'}(\cdot) = \bmphi^T(\cdot) \,  \hat{\bmxi}_{jj'}$, where
\eq
\hat{\bmxi}_{jj'} := (\bB^T \bB + \lam \bOm)^{-1} \bB^T \hat{\bmrho}_{jj'}
\eqq
and $\hat{\bmrho}_{jj'} := [\tilde{\sig}_{jj'}(\al_1),\dots,\tilde{\sig}_{jj'}(\al_L)]^T$.
\item[(b)] As $n \linfty$, if\/ $\hat{\bGam}(\tau,\al) \plim \bGam(\tau,\al)$ for any fixed $\tau$ and $\al$,
then $\hat{\bV}(\al)  \plim \bar{\bV}(\al) := [\bar{\sig}_{jj'}(\al)]_{j,j'=1}^m$ 
uniformly in $\al \in [\alu,\albar]$, where $\bar{\sig}_{jj'}(\al) := \bmphi^T(\al) \, \bar{\bmxi}_{jj'}$ and 
$\bar{\bmxi}_{jj'} := (\bB^T \bB + \lam \bOm)^{-1} \bB^T \bar{\bmrho}_{jj'}$,
with $\bar{\bmrho}_{jj'}$ being the vector made of the $(j,j')$-th entry of
\eq
\bSig(\al_\ell)  :=  \bGam(0,\al_\ell) - \, \bar{\bA}(\al_\ell) \, \bmgam_p^T(\al_\ell)  
- \bmgam_p(\al_\ell)   \, \bar{\bA}^T(\al_\ell) + \bar{\bA}(\al_\ell) \, \bGam_p(\al_\ell) \, \bar{\bA}^T(\al_\ell)
\eqq
for $\ell=1,\dots,L$.
\item[(c)]  If $\bA_{p,\tau}(\cdot) \in \cF_m$ for all $\tau=1,\dots,p$ and $\bV_p(\cdot) \in \cF_m$, 
then,  as $\lam \lzero$, 
$\bar{\bV}_p(\al) \rightarrow \bV_p(\al)$ uniformly in $\al \in [\alu,\albar]$.
\end{itemize}
}
\label{thm:v}
\end{pro}

\noindent
{\sc Proof}: Assertion (a) is the standard result from spline smoothing (Hastie and Tibshirani 1990, p.\ 28).
To prove assertion (b), observe that $\tilde{\bV}(\al_\ell)  := [\tilde{\sig}_{jj'}(\al_\ell)]_{j.j'=1}^m 
= (n-p)^{-1} \| \bY_\ell - \hat{\bThe} \bZ_\ell \|^2$. Combining this expression with Proposition~\ref{thm:a}(b) 
yields $\tilde{\bV}(\al_\ell) \plim \bSig(\al_\ell)$. This implies that $\hat{\bmrho}_{jj'} \plim \bar{\bmrho}_{jj'}$.
Hence the assertion (b). Under the condition of (c), $\bar{\bA}(\al_\ell) \rightarrow \bA_p(\al_\ell)
= \bmgam_p(\al_\ell) \, \bGam_p^{-1}(\al_\ell)$ by Proposition~\ref{thm:a}(b). Therefore, $\bSig(\al_\ell) \rightarrow \bGam(0,\al_\ell) - \, \bA_p(\al_\ell) \, \bmgam_p^T(\al_\ell) = \bV_p(\al_\ell)$.
This implies that $\bar{\bmrho}_{jj'} \rightarrow \bmrho_{jj'}$, where $\bmrho_{jj'}$ is the vector made
of the $(j,j;)$-th entry of $\bV_p(\al_\ell)$ for $\ell=1,\dots,L$. If $\bV_p(\cdot) \in \cF_m$, then
$\sig_{jj'}(\al) = \bmphi^T(\al)$ for some $\bmxi_{jj'} \in \bbR^K$. This implies that 
$\bmrho_{jj'} = \bB \, \bmxi_{jj'}$. In this case, $\bar{\bmxi}_{jj'} \rightarrow \bmxi_{jj'}$ and hence
$\bar{\sig}_{jj'}(\al) \rightarrow \bmphi^T(\al) \, \bmxi_{jj'} = \sig_{jj'}(\al)$  uniformly in $\al \in [\alu,\albar]$.
\qed

\medskip
By following  the way in which  $\bS_p(\om,\al)$ is defined by the Yule-Walker solutions $\bA_p(\cdot)$ and $\bV_p(\cdot)$, let $\bar{\bS}_p(\om,\al)$ denote the  corresponding spectrum defined by $\bar{\bA}_p(\cdot)$ and $\bar{\bV}_p(\cdot)$ from Propositions~\ref{thm:a} and \ref{thm:v}.
In light of the above analysis, $\bar{\bS}_p(\om,\al)$ 
can be viewed as a regularized version of $\bS_p(\om,\al)$ based on smoothing spline parameters in $\cF_m$. 
As an immediate result of Propositions~\ref{thm:a} and \ref{thm:v}, the following theorem summarizes the relations between these spectra and the SAR estimator $\hat{\bS}(\om,\al)$ in (\ref{Shat}).

\medskip
\noindent
{\bf Theorem}. If $\hat{\bGam}(\tau,\al) \plim \bGam(\tau,\al)$ as $n \linfty$ for fixed $\tau$ and $\al$, then $\hat{\bS}(\om,\al) \plim \bar{\bS}_p(\om,\al)$ uniformly in $(\om,\al) \in [0,2 \pi) \times [\alu,\albar]$. 
In addition, if the AR parameters in (\ref{AR}) are members of $\cF_m$, then 
$\bar{\bS}_p(\om,\al) \rightarrow \bS_p(\om,\al)$ as $\lam \lzero$ uniformly 
 in $(\om,\al) \in [0,2 \pi) \times [\alu,\albar]$.

\section{ Simulation Study}

To evaluate the SAR estimator, we present the results of a simulation study 
using a set of simulated data with $m=2$. Additional results can be found in  Appendix III. 

Let $\{ \xi_{1,t} \}$, $\{ \xi_{2,t} \}$, and $\{ \xi_{3,t} \}$ be 
zero-mean and unit-variance AR series satisfying $\xi_{1,t} =  a_{11} \, \xi_{1,t-1} + \ep_{1,t}$,
$\xi_{2,t} =  a_{21} \, \xi_{2,t-1} + \ep_{2,t}$, and $\xi_{3,t} =  a_{31} \, \xi_{3,t-1} + a_{32} \, \xi_{3,t-2} + \ep_{3,t}$,
where $a_{11}  := 0.8$, $a_{21} := -0.7$, $a_{31} := 2 d \cos(2\pi f_0)$, and 
$a_{32} := -d^2$ with $d=0.9$ and $f_0 =0.2$, and where $\{\ep_{1,t} \}$, $\{\ep_{2,t} \}$, 
and  $\{\ep_{3,t} \}$  are mutually independent Gaussian white noise. By construction, 
the conventional spectrum of $\{ \xi_{1,t} \}$ has a broad peak at $\om = 0$, the conventional spectrum 
of $\{ \xi_{2,t} \}$ has a broad peak at $\om = \pi$, and the conventional spectrum
of $\{ \xi_{3,t} \}$ has a narrow peak at $\om = 2 \pi \times 0.2$.

Let $\{ z_{t} \}$ be a nonlinear mixture
of $\{ \xi_{1,t} \}$ and $\{ \xi_{2,t} \}$ defined by
\eq
z_{t}   :=  \psi_1(\xi_{1,t}) \times \xi_{1,t}+ (1-\psi_1(\xi_{1,t})) \times \xi_{2,t},
\eqq
where $\psi_1(y) := 0.9 I(y < -0.8) + 0.2  I(y > 0.8) + \{ 0.9 - (7/16) (y + 0.8)\} I(|y| \le 0.8)$.
Because $\psi_1(y)$ equals 0.9 for $y < -0.8$ and 0.2 for $y > 0.8$, the series
$\{ z_t \}$ behaves very similarly to $\{ \xi_{1,t} \}$ at lower quantiles and somewhat  similarly 
to $\{ \xi_{2,t} \}$ at higher quantiles. The final series $\by_t := [y_{1,t}, y_{2,t}]^T$   is given by
\eqn
 \left\{
\begin{array}{l}
y_{1,t}  :=  \psi_2(z_{t}) \times z_{t} + (1-\psi_2(z_{t})) \times \xi_{3,t}, \\
y_{2.t}  :=  \xi_{3,t+10},
\end{array} \right.
\label{y}
\eqqn
where $\psi_2(y) := 0.5 I(y <  -0.4) +  I(u > 0.4) + \{ 0.5 + (5/8) (u + 0.4)\} I(|u| \le 0.4)$.
Because $\psi_2(y)$ equals 0.5 for $y < -0.4$ and 1 for $y > 0.4$,  the series $\{y_{1,t}\}$ behaves 
similarly to $\{ z_t\}$ at higher quantiles and blends the characteristics of $\{ z_t\}$ and $\{ \xi_{3,t} \}$ 
at lower quantiles. The series $\{ y_{2,t}\}$ is a copy of $\{ \xi_{3,t} \}$ delayed by 10 units of time.
It is conceivable that the summability requirement for $r_{jj'}(\tau, \alpha)$  should 
be satisfied by the process in (\ref{y}) due to the strong mixing property 
of each component with fast vanishing mixing numbers (Mokkadem 1988).

\begin{figure}[p]
\centering
\includegraphics[height=4in,angle=-90]{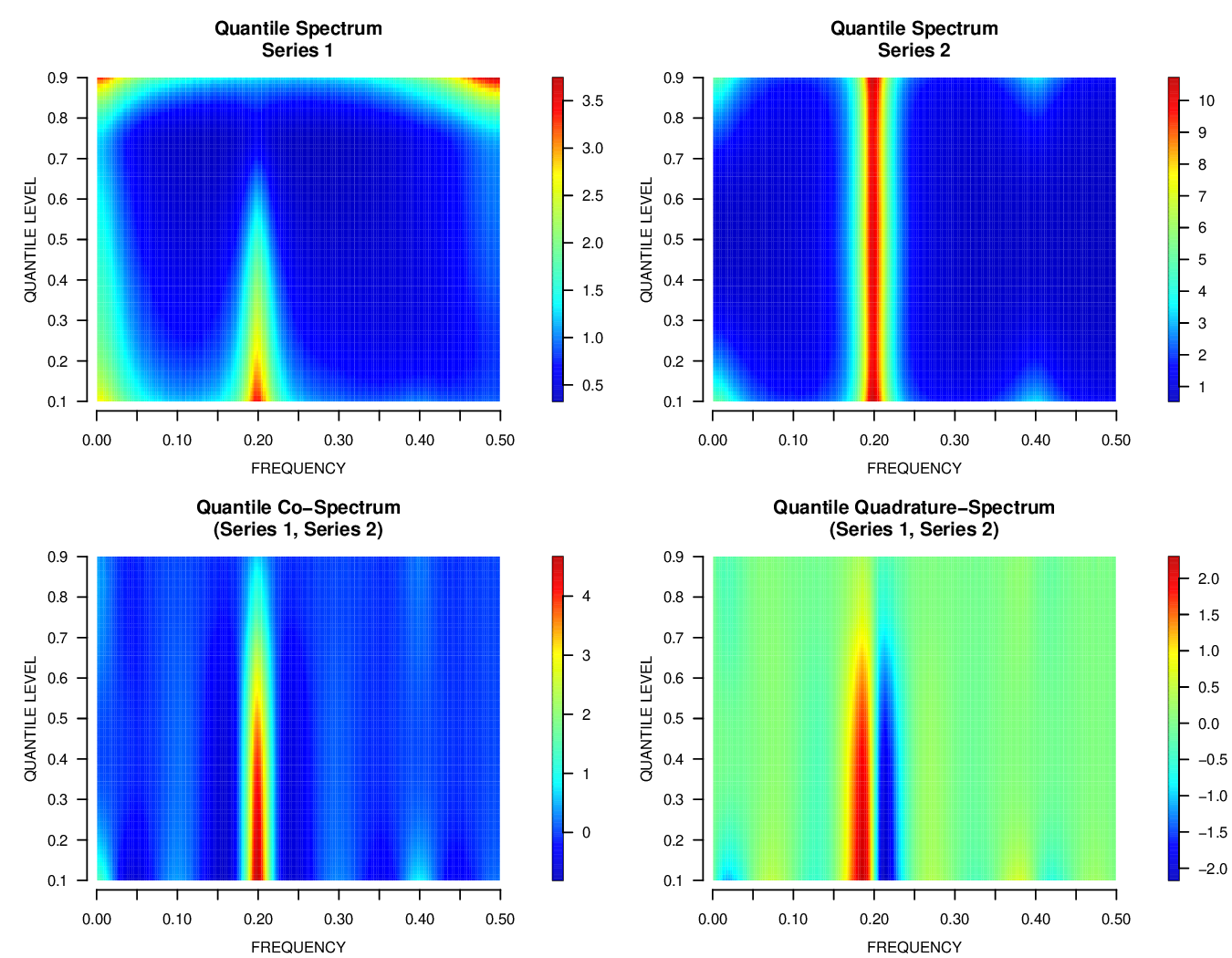}  
\caption{Quantile spectrum of the mixture process (\ref{y}).  
Real and complex parts of the cross-spectrum (second row) are 
known as co-spectrum and quadrature-spectrum, respectively. All spectra are shown as functions of frequency
variable $f := \om/(2\pi) \in (0,0.5)$. }
 \label{fig:qspec}
 \centering
 \vspace{0.2in}
 \includegraphics[height=4in,angle=-90]{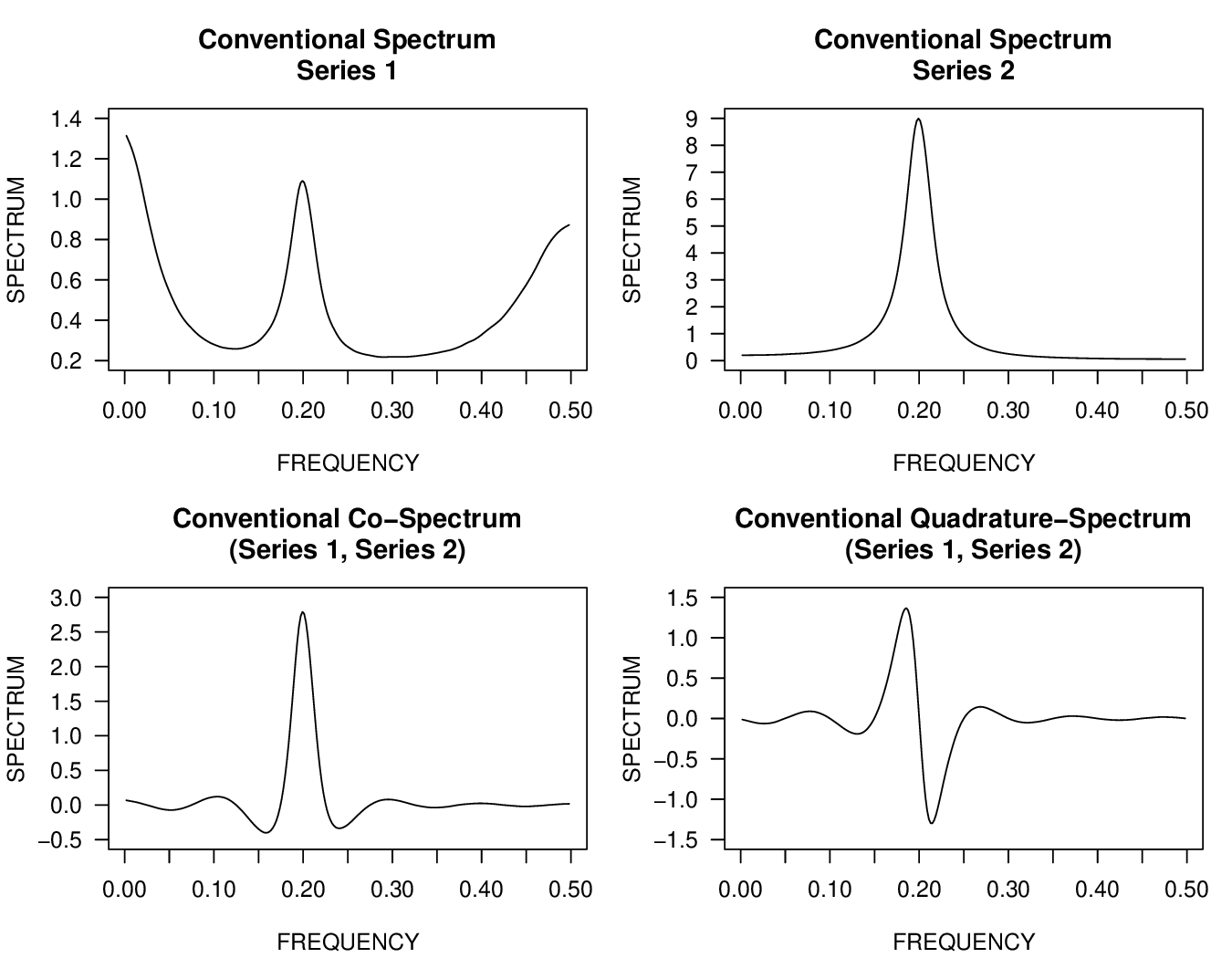}  
\caption{Conventional spectrum of the mixture process (\ref{y}). }
 \label{fig:spec}
\end{figure}

Figure~\ref{fig:qspec} depicts the quantile spectrum of the process defined by (\ref{y}).
This is the ensemble mean of the quantile periodograms 
defined by (\ref{qper}) from 5000  Monte Carlo runs, computed at $\om_v = 2\pi  v/n$ 
$(v=1,\dots,\lfloor (n-1)/2 \rfloor; n=512)$ and $\al_\ell = 0.1 + 0.01 (\ell-1)$ $(\ell=1,\dots,81)$.  
Extreme quantiles are excluded due to their different statistical properties 
(Koenker 2005, p.\ 130; Davis and Mikosch 2009). The granularity of quantile levels is chosen somewhat 
arbitrarily (sufficiently small to capture the changing patterns).
Compared to the conventional spectrum shown in Figure~\ref{fig:spec},
the quantile spectrum in Figure~\ref{fig:qspec}  reveals the 
quantile-dependent characteristics of  $\{ y_{1,t}\}$. The quantile co-spectrum 
and quadrature-spectrum  reveal a strong 
correlation between $\{y_{1,t}\}$ and $\{y_{2,t}\}$ in the lower and middle quantile 
region around frequency $2\pi \times 0.2$.

\begin{figure}[p]
\centering
\includegraphics[height=4in,angle=-90]{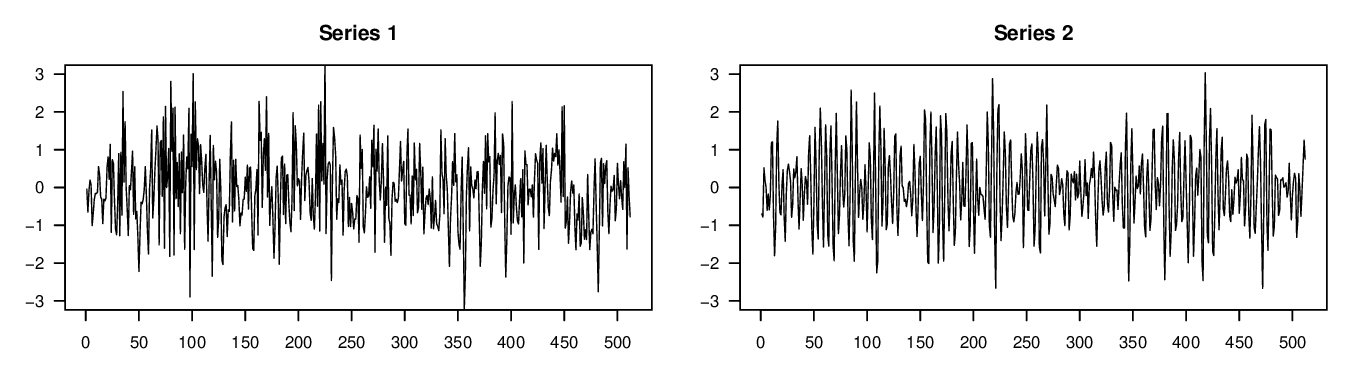} 
\caption{An example of simulated time series $(n=512)$ according to (\ref{y}). }
\label{fig:ts}
\vspace{0.2in}
\includegraphics[height=4in,angle=-90]{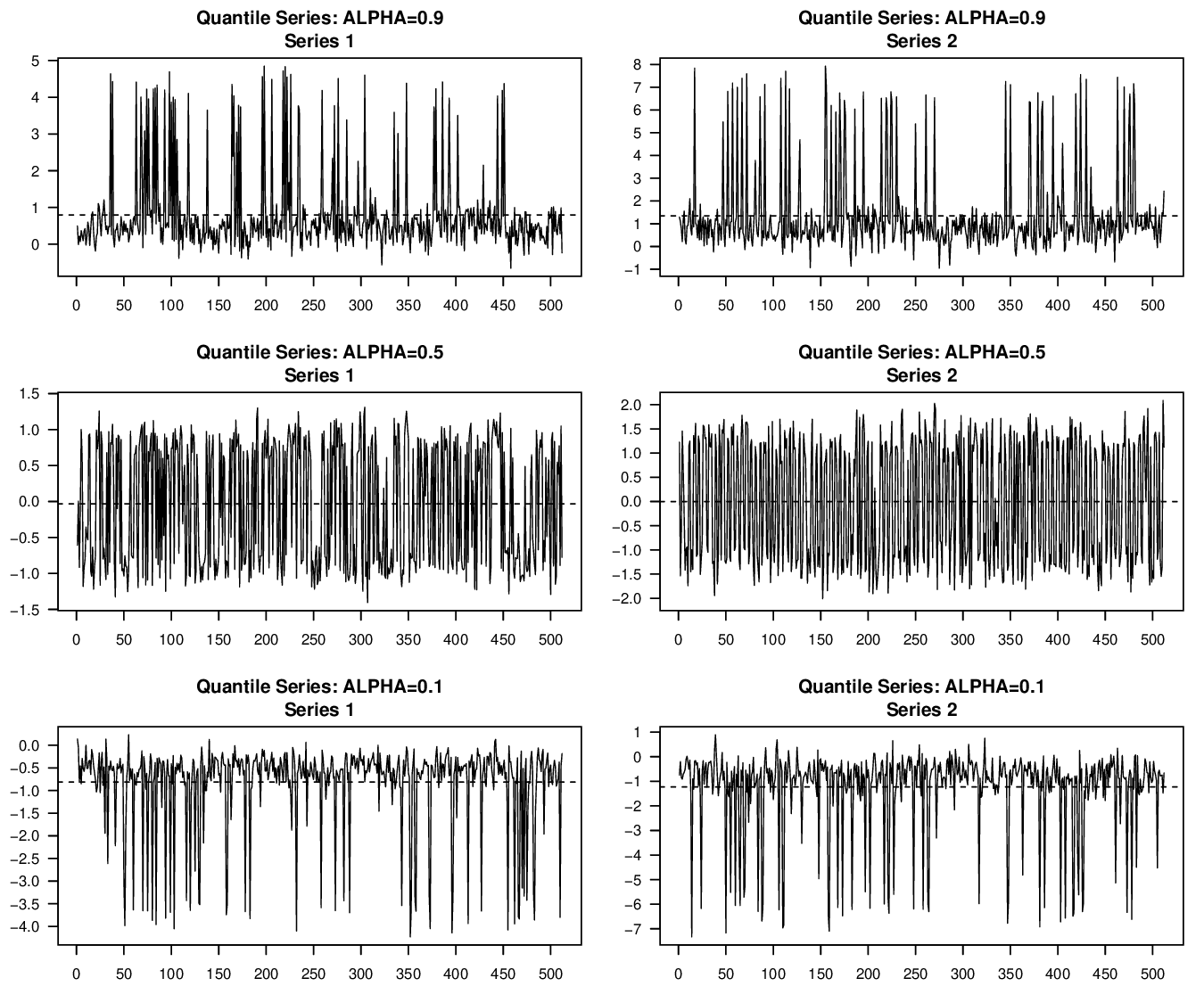} 
\caption{Quantile series obtained from the series shown in Figure~\ref{fig:ts}
at $\al =0.9$ (first row), $\al =0.5$ (second row), and $\al=0.1$ (third row).  
Dashed horizontal line depicts the sample mean of quantile series 
which coincides with the sample quantile of the original series. }
\label{fig:qser}
\end{figure}

\begin{figure}[t]
\centering
\includegraphics[height=4in,angle=-90]{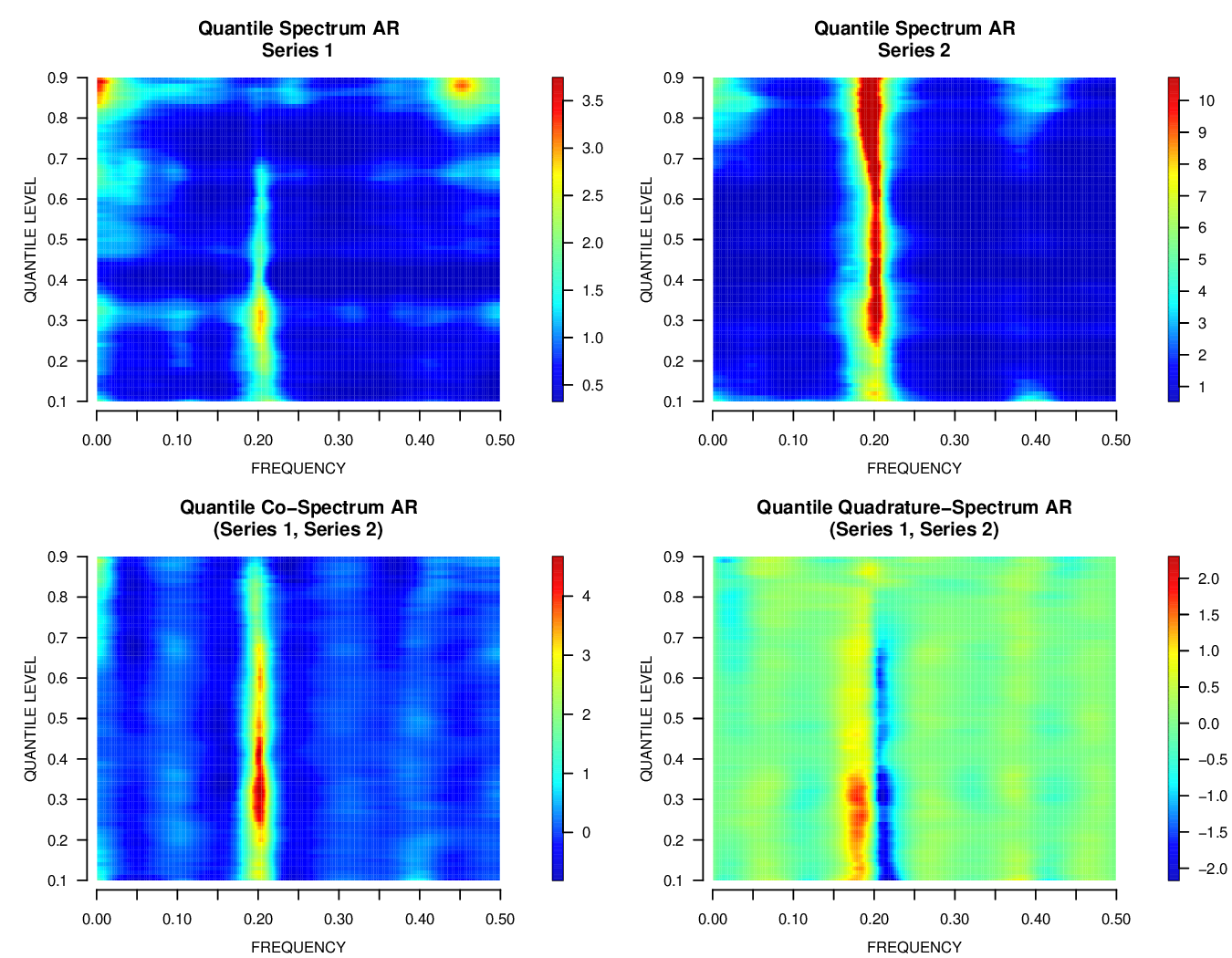}  
\caption{The AR estimate (without quantile smoothing) 
of the quantile spectrum shown in Figure~\ref{fig:qspec} obtained 
from the series shown in Figure~\ref{fig:ts}. (KLD = 0.183).}
 \label{fig:ar}
\end{figure}

Figure~\ref{fig:ts} shows an example of the time series generated according to (\ref{y}). 
Figure~\ref{fig:qser} depicts the corresponding QSER sat quantile levels $\al=0.1$, $0.5$, and $0.9$. 
Figure~\ref{fig:ar} shows the spectral estimate obtained from this series by the AR estimator 
without quantile smoothing.
This estimate is able to capture some key features of the underlying 
spectrum, including the spectral peak around frequency $2 \pi \times 0.2$ 
and the dependency of its magnitude on quantile levels.
However, without quantile smoothing, this estimate remains noisy across quantiles.

To measure the accuracy of spectral estimation, we employ  the Kullback-Leibler spectral divergence defined by
\eq
{\rm KLD} := 
\frac{1}{L \lfloor (n-1)/2 \rfloor} \sum_{\ell=1}^L
\sum_{v=1}^{\lfloor (n-1)/2 \rfloor} \bigg\{ \tr \big( \hat{\bS}(\om_v,\al_\ell) \, \bS^{-1}(\om_v,\al_\ell) \big)
- \log \frac{| \hat{\bS}(\om_v,\al_\ell)| }{ |\bS(\om_v,\al_\ell)| } - m \bigg\}.
\eqq
This is a nonegative quantity which equals zero when $\hat{\bS}(\om_v,\al_\ell) = \bS(\om_v,\al_\ell)$
for all $v$ and $\ell$. The KLD is related to Whittle's likelihood for time series modeling (Whittle 1953) and has 
been used as the dissimilarity measure of conventional spectra for time series clustering and classification (Kakizawa et al.\  1998). 
For the AR estimate shown in Figure~\ref{fig:ar}, we have KLD = 0.183.

\begin{figure}[H]
\centering
\includegraphics[height=3.2in,angle=-90]{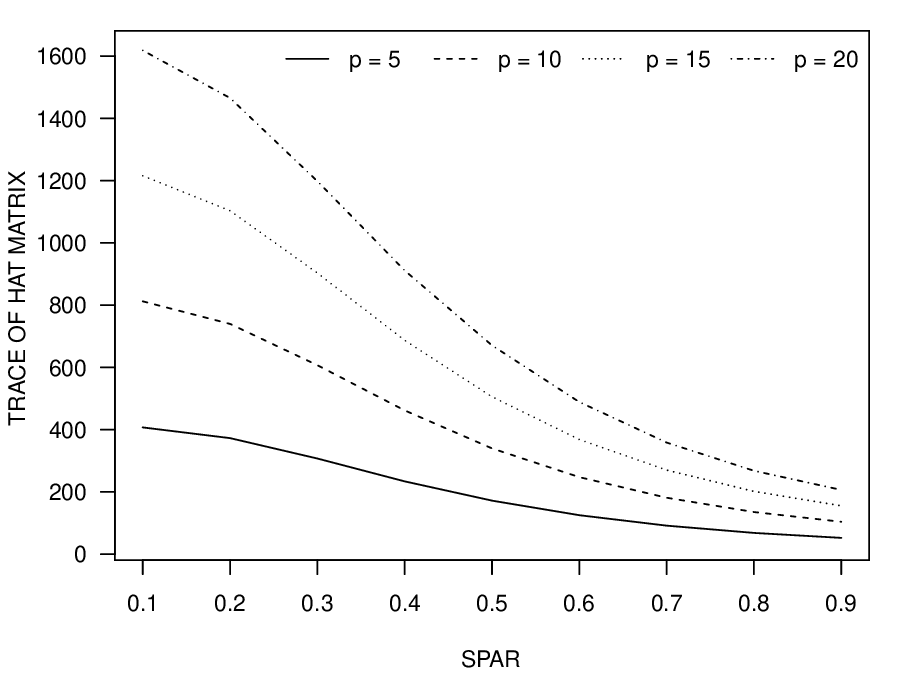}  
\caption{Illustration of $\tr(\bH)$ in the SAR problem (\ref{sar}) as a function of the smoothing parameter {\tt spar} for different values of the order parameter $p$.}
\label{fig:df}
\end{figure}

Before presenting the SAR estimate, we  would like to use Figure~\ref{fig:df} to demonstrate the difficulty of selecting $p$ and $\lam$ jointly using $\tr(\bH)$ as the effective degree of freedom. 
In this figure, $\tr(\bH)$  is plotted against the smoothing parameter {\tt spar}, a reparameterized monotone function of $\lam$ (see Appendix I), for some fixed values of $p$. As expected for a meaningful degree-of-freedom measure, 
the trace decreases with {\tt spar} for fixed $p$ and increases with $p$ for fixed {\tt spar}. 
However, when considered jointly, the effect of a large $p$ can be mitigated by the choice of a large {\tt spar}, resulting in an unchanged value of $\tr(\bH)$.

\begin{figure}[t]
\centering
\includegraphics[height=4in,angle=-90]{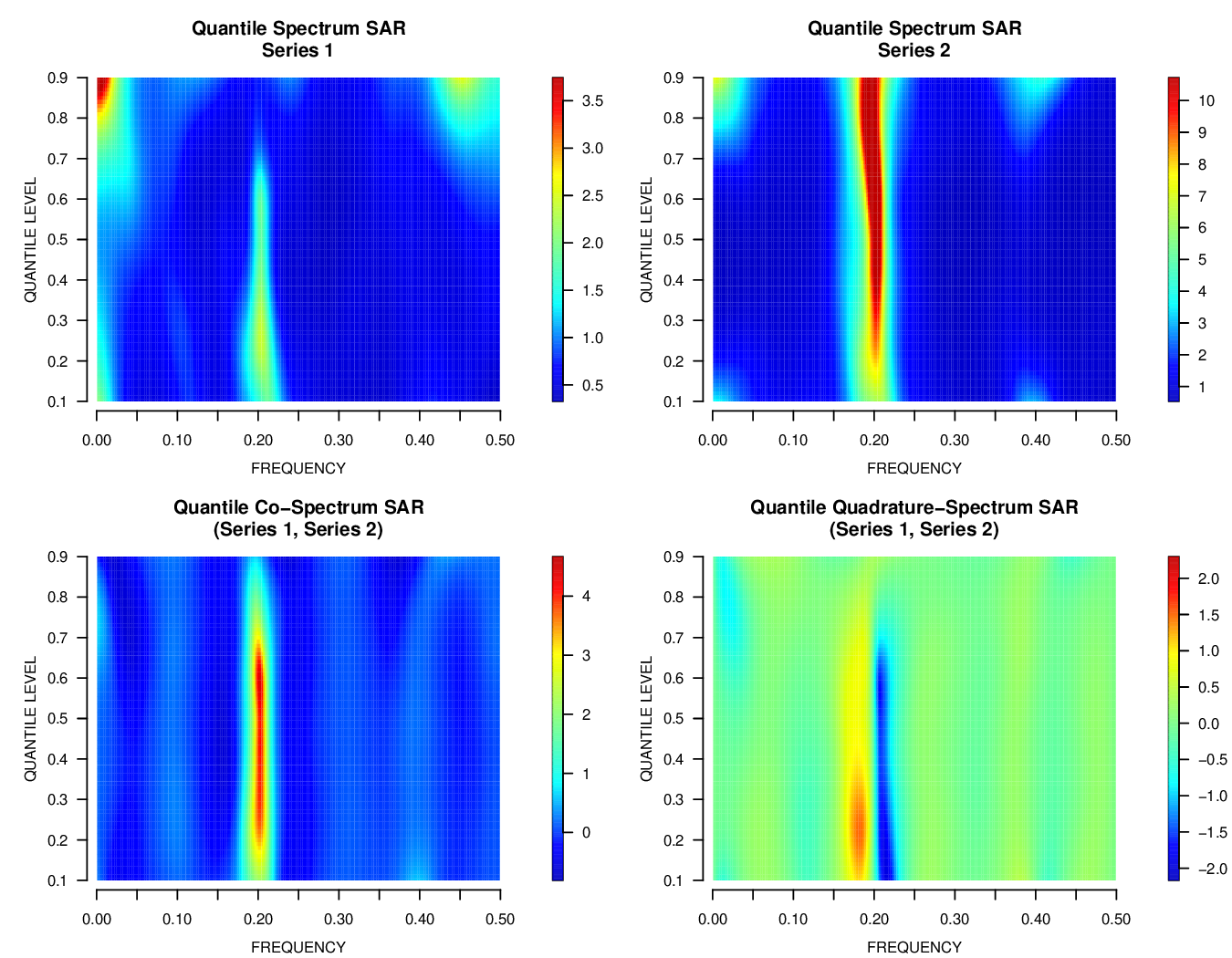}    
\caption{The SAR estimate of the quantile spectrum shown in Figure~\ref{fig:qspec} from the series shown in Figure~\ref{fig:ts} with the smoothing parameter selected by GCV.  (KLD = 0.100, {\tt spar} = 0.904).}
\label{fig:sargcv}
\end{figure}

Returning to spectral estimation,  Figure~\ref{fig:sargcv} shows the SAR estimate obtained from 
the series in Figure~\ref{fig:ts} with {\tt spar} determined by the GCV in (\ref{gcv}). This estimate exhibits remarkably improved smoothness across quantiles in comparison with the estimate shown in Figure~\ref{fig:ar}. The improvement in appearance is reflected in the reduced KLD, which equals 0.100.

\begin{table}[H]
\vspace{0.2in}
{\footnotesize
\begin{center} 
\caption{Mean KLD  of Spectral Estimators for the Mixture Process  (\ref{y})}
\label{tab:err}
\begin{tabular}{ccccccccccccc} \toprule
&  \multicolumn{2}{c}{SAR} &  \multicolumn{3}{c}{AR} &  \multicolumn{3}{c}{LW} \\
 $n$   &       GCV     & Fixed {\tt spar}  &  None & SPLINE  & GAMM   & None & SPLINE  & GAMM \\   \midrule
 256  &  0.194   & 0.181  &  0.309 & 0.303 & 0.230  & 0.313 & 0.307 & 0.233 \\ 
  512 &  0.098  & 0.097  &  0.178  & 0.175  & 0.119  & 0.204 & 0.200 & 0.138 \\
\hline
\end{tabular} 
\end{center}
}
{\scriptsize 
\begin{center}
\begin{minipage}{5in}
Results are based on 1000 Monte Carlo runs. Fixed {\tt spar}:   {\tt spar} = 1 for $n=256$ and 
{\tt spar} = 0.9 for $n = 512$.
SPLINE:  {\tt smooth.spline}. GAMM: {\tt gamm} with correlated residuals. 
LW: lag-window estimator using Tukey-Hanning window with optimal bandwidth parameter 
($M=24$ for $n=256$ and $M=30$ for $n=512$). 
\end{minipage}
\end{center}
}
\end{table}

\begin{figure}[t]
\centering
\vspace{0.2in}
\includegraphics[height=3in,angle=-90]{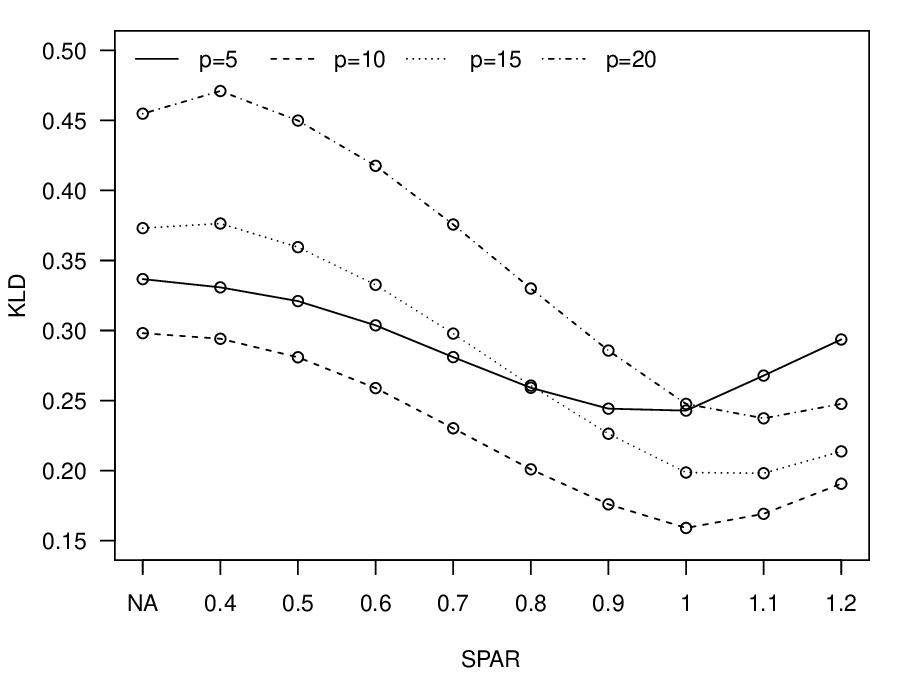}
\includegraphics[height=3in,angle=-90]{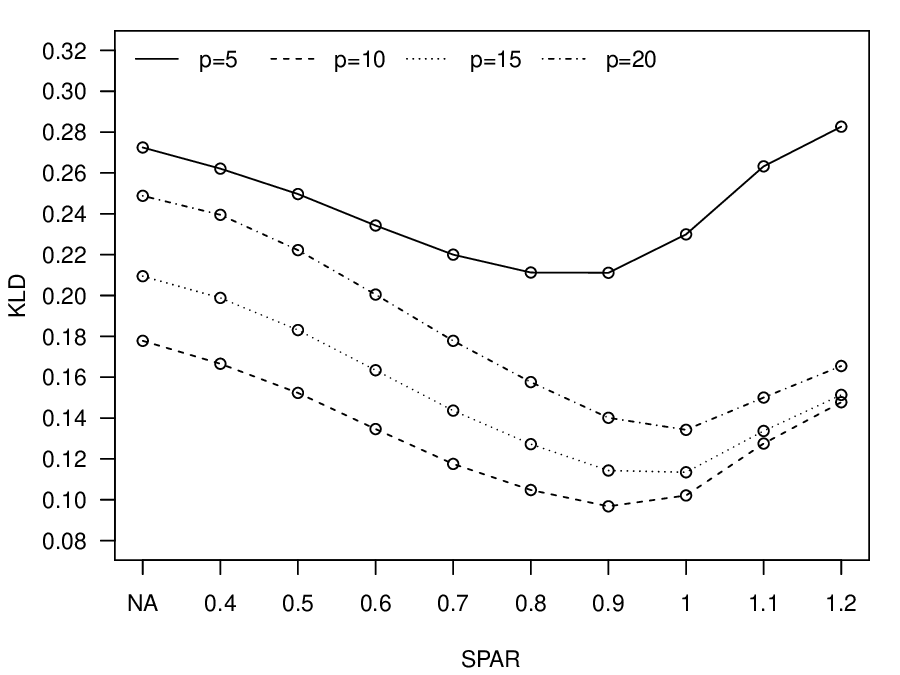}
\centerline{\hspace{0.3in}(a) \hspace{2.8in}(b)}
\caption{Mean KLD of the SAR estimator for the mixture process (\ref{y}) with different choices of $p$ and {\tt spar} 
({\tt spar} = NA corresponds to the AR estimator without quantile smoothing). (a) $n=256$. (b) $n=512$.
Results are based on 1000 Monte Carlo runs.}
\label{fig:err}
\end{figure}

A more comprehensive comparison is presented in Table~\ref{tab:err}. This table contains 
the mean KLD of the SAR estimator computed from 1000 Monte Carlo runs with two 
sample sizes. The fixed value of {\tt spar} corresponds to the minimizer
of the mean KLD when $p=10$, as shown in Figure~\ref{fig:err}.  The order $p= 10$ 
yields the best result against the other choices in Figure~\ref{fig:err}. This can be explained by
the fact that $\{ y_{2,t} \}$ is a copy of $\{ \xi_{3,t} \}$ delayed  by 10 units of time. 
According to Figure~\ref{fig:err}, there exists a range of choices for {\tt spar} to yield an improved KLD
 for the SAR estimator over the AR estimator without quantile smoothing.

Besides the SAR estimator, 
Table~\ref{tab:err} also contains the results from the AR and LW estimators with and 
without quantile smoothing. Two procedures are employed for quantile smoothing: the R functions {\tt smooth.spline} and {\tt gamm}.  While {\tt smooth.spline} performs simple spline smoothing with GCV
(Hastie and Tibshirani 1990, p.\ 27), {\tt gamm}  from the package `mgcv'  (Wood 2022) 
incorporates correlated residuals in the form a random effect with AR(1)-type correlation under the framework 
of generalized additive mixed-effect model. Table~\ref{tab:err} shows the effectiveness of {\tt gamm} 
in comparison with {\tt smooth.spline} for quantile smoothing in both AR and LW estimators. These 
estimators are outperformed by the SAR estimator in Table~\ref{tab:err}.

A benefit of the SAR method is that the time-domain AR model with quantile-dependent 
functional coefficients can be used to perform Granger-causality analysis 
(Granger 1963; L\"{u}tkepohl 1993, p.\ 93) across quantiles.
Details of this method can be found in Appendix II.

\begin{figure}[p]
\centering
\includegraphics[height=6.2in,angle=-90]{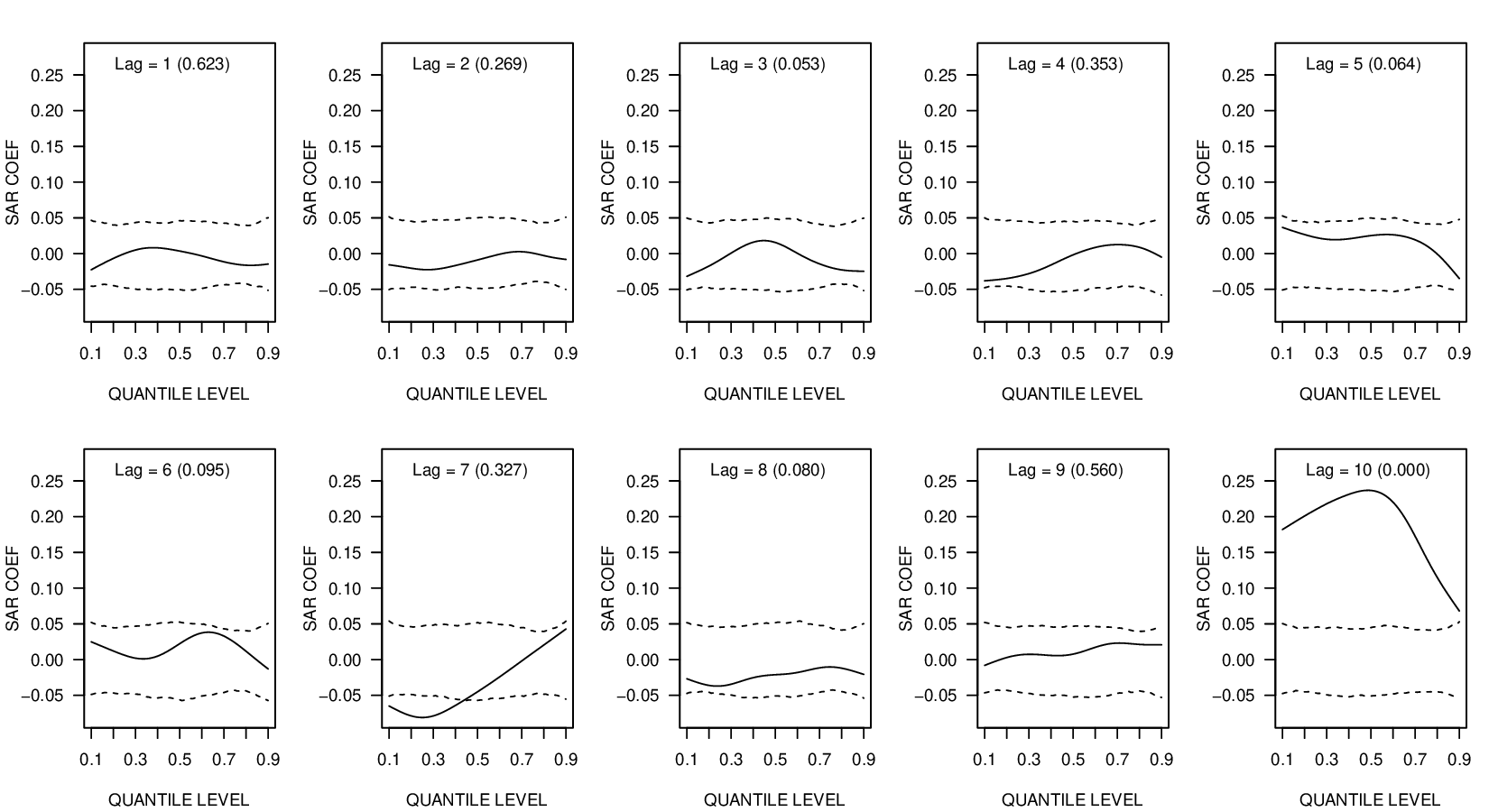}  
\caption{The (1,2)-entry of $\hat{\bA}_\tau(\cdot)$ at $\tau=1,\dots,10$
for the series shown in Figure~\ref{fig:ts}.
Dashed lines depict the pointwise 95\% bootstrap confidence band constructed from 
1000 bootstrap samples under the null hypothesis of no Granger-causality.
Numbers in parentheses are $p$-values of the bootstrap Wald statistic.}
\label{fig:causality}
\vspace{0.5in}
\begin{table}[H]
{\footnotesize
\begin{center} 
\caption{Mean $p$-Values of Wald Test on  (1,2)-Entry 
for the Mixture Process (\ref{y})}
\label{tab:causality}
\begin{tabular}{lccccccccccccc} \toprule
 $\tau=1$ & $\tau=2$ & $\tau=3$ & $\tau=4$ & $\tau=5$ & $\tau=6$ & $\tau=7$ & $\tau=8$ & $\tau=9$ & $\tau=10$ & All $\tau$ \\  \midrule
  0.329& 0.331& 0.340& 0.293& 0.172& 0.310& 0.188& 0.184& 0.199& {\bf 0.000} & {\bf 0.000} \\
\hline
\end{tabular} 
\end{center}
}
{\scriptsize 
\begin{center}
\begin{minipage}{5.2in}
Results are based on 1000 Monte Carlo run ($n=512)$. The $p$-value in each run is computed from 1000 bootstrap samples.  Boldface font highlights the cases where $p$-value is less than 0.05.
\end{minipage}
\end{center}
}
\end{table}
\end{figure}

According to  (\ref{y}), the lagged series $\{ y_{2,t-\tau}\}$ is expected to exhibit a strong 
effect of Granger-causality for the series $\{ y_{1,t}\}$ at $\tau = 10$. This expectation is confirmed 
by the SAR-based Granger-causality analysis shown in Figure~\ref{fig:causality}. 
The (1,2)-entry of $\hat{\bA}_\tau(\cdot)$ at lag $\tau=10$ as a function of $\al$
resides entirely outside the 95\% bootstrap confidence band constructed under the null hypothesis
of no Granger-causality. This entry at other lags lies mostly inside the respective confidence band. 
The statistical significance of this Granger-causality  is  manifested 
 in the $p$-value of 0.000 for the corresponding bootstrap Wald statistic.

Table~\ref{tab:causality} contains the result of the bootstrap Wald test from 
1000 Monte Carlo runs. As can be seen, this SAR-based test  is able to detect the Granger-causality at $\tau=10$ with mean $p$-value equal to 0.000, whereas at the remaining lags the mean 
$p$-value is no less than 0.172.

\section{Concluding Remarks}

This paper proposes the spline autoregression (SAR) method for estimating the quantile spectrum  introduced in Li (2008; 2012). The SAR spectral estimator is a bivariate function of frequency and quantile level. It is derived from an autoregression model where the coefficients  are spline functions of the quantile level. This model is  fitted by penalized least-squares to a set of quantile series (QSER) to produce a smoothing spline estimate. 
The simulation study validates the proposed method as an effective way of leveraging the smoothness
of the quantile spectrum with respect to the quantile level to produce more accurate spectral estimates.

A key enabler of the SAR method is the creation of the QSER as  the inverse Fourier transform of the quantile discrete Fourier transform (QDFT) computed by trigonometric quantile regression.
It is conceivable that this machinery could be applied to the result of trigonometric $M$-regression
where the objective function $\rho_\al(\cdot)$ is replaced by another nonnegative function, as in Fajardo et al.\ (2018), which may be further indexed by a continuous parameter analogous to $\al$. Instead of periodogram smoothing, an AR or SAR model could be fitted to the resulting counterpart of the QSER to produce 
an estimate of the corresponding spectrum.

The SAR method can be extended to provide an estimator of the conventional spectrum
of the level-crossing processes $\{ I(y_{j,t} \le q_j(\al)) \}$ $(j=1,\dots,m)$ (or quantile-crossing spectrum) 
as a bivariate function of frequency and quantile level. This problem is addressed in Li (2024) as a sequel
to the current paper.

For conventional spectral analysis, the ARMA model is a more flexible extension of the AR model (Percival and Walden 1993; Stoica and Moses 1997), which is especially effective for time series with deep spectral troughs. An interesting topic for future research is the employment of the ARMA model to estimate the quantile spectrum.

\section*{References}

{\footnotesize
\begin{description}

\item
Barun\'{i}k, J., and Kley, T. (2019). Quantile coherency: A general measure for dependence between cyclical economic variables. {\it Econometrics Journal}, 22, 131–152.

\item
Birr, S., Volgushev, S., Kley, T., Dette, H., and Hallin, M. (2017). Quantile spectral
analysis for locally stationary time series. {\it Journal of the Royal Statistical Society Series B}, 79, 1619--1643.

\item
Birr, S., Kley, T., and Volgushev, S. (2019). Model assessment for time series dynamics
using copula spectral densities: A graphical tool. {\it Journal of Multivariate Analysis}, 172, 122--146.

\item Brockwell, P., and Davis, R. (1991). {\it Time Series: Theory and Methods}, 2nd edn. New York: Springer-Verlag.

\item
Chen, T., Sun, Y., and Li, T.-H. (2021). A semi-parametric estimation method for the quantile
spectrum with an application to earthquake classification using convolutional  neural network. {\it Computational Statistics and Data Analysis}, 154, 107069.

\item
Cheng, H., Wang, Y., Wang, Y., and Yang, T. (2022). Inferring causal interactions in financial markets using conditional Granger causality based on quantile regression. {\it Computational Economics}, 59, 719--748.

\item
Choi, B. (1993). Multivariate maximum entropy spectrum. {\it Journal of Multivariate Analysis}, 46, 56--60.

\item
Chuang, C.-C., Kuan, C.-M., and Lin, H.-Y. (2009).
Causality in quantiles and dynamic stock return–volume relations. {\it Journal of Banking and Finance}, 33, 1351--1360.

\item
Davis, R., and Mikosch, T. (2009). The extremogram: A correlogram for extreme events. 
{\it Bernoulli}, 15, 977--1009.

\item 
Dette, H., Hallin, M., Kley, T., and Volgushev, S. (2015). Of copulas,
quantiles, ranks and spectra: an $L_1$-approach to spectral analysis. 
{\it Bernoulli}, 21, 781--831.

\item
Fajardo, H., Reisen, V., L\'{e}vy-Leduc, C.,  and Taqqu, M. (2018). $M$-periodogram for 
the analysis of long-range dependent time series. {\it Statistics}, 52, 665--683.

\item
Goto, Y., Kley, T., Van Hecke, R., Volgushev, S., Dette, H., and Hallin, M. (2022). The
integrated copula spectrum. {\it Annals of Statistics}, 50, 3563--3591.

\item
Granger, C. (1969). Causal relations by econometric models and cross-spectral methods. {\it Econometrica}, 37, 424-438.

\item
Hagemann, A. (2013). Robust spectral analysis. arXiv:1111.1965.

\item
Hastie, T., and Tibshirani, R. (1990). {\it Generalized Additive Models}. Boca Raton, FL: CRC Press.

\item
Hong, Y. (2000). Generalized spectral tests for serial dependence. {\it Journal of the Royal Statistical Society Series B}, 62, 557--574.

\item
Jim\'{e}nez-Var\'{o}n, C., Sun, Y., and Li, T.-H. (2024).
A semi-parametric estimation method for  quantile coherence  with an application to bivariate financial time series clustering. 
{\it Econometrics and Statistics}. \url{https://doi.org/10.1016/j.ecosta.2024.11.002}

\item
Jordanger, L., and Tj{\o}stheim, D. (2022). Nonlinear spectral analysis: a local
Gaussian approach. {\it  Journal of the American Statistical Association}, 117, 1010--1027. 

\item
Jordanger, L., and Tj{\o}stheim, D. (2023). Local Gaussian cross-spectrum analysis.
{\it Econometrics}, 11, 1--27.

\item
Kakizawa, Y., Shumway, R., and Taniguchi, M. (1998). Discrimination and clustering for multivariate time series. {\it  Journal of the American Statistical Association}, 93, 328--340.

\item
Kedem, B. (1986). Spectral analysis and discrimination by zero-crossings. {\it Proceedings of the IEEE},  74, 1477--1493.

\item
Koenker, R. (2005). {\it Quantile Regression}. Cambridge, UK: Cambridge University Press.

\item
Lahiri, S.\ (2003). {\it Resampling Methods for Dependent Data}, New York: Springer.

\item
Li, T.-H. (2008) Laplace periodogram for time series analysis. {\it Journal of the American Statistical Association}, 103, 757--768.

\item
Li, T.-H. (2012). Quantile periodograms. {\it Journal of the American Statistical Association}, 107, 765--776. 

\item 
Li, T.-H. (2014a). {\it Time Series with Mixed Spectra}. Boca Raton, FL: CRC Press.

\item
Li, T.-H. (2014b). Quantile periodogram and time-dependent variance.
{\it Journal of Time Series Analysis}, 35, 322--340.

\item
Li, T.-H. (2020). From zero crossings to quantile-frequency analysis of time
series with an application to nondestructive evaluation. 
{\it Applied Stochastic Models for Business and Industry}, 36, 1111-1130.

\item
Li, T.-H. (2021). Quantile-frequency analysis and spectral measures
for diagnostic checks of time series with nonlinear
dynamics. {\it Journal of the Royal Statistical Society Series C},  70, 270--290.

\item
Li, T.-H. (2023), Quantile-frequency analysis and deep learning for signal classification.
{\it Journal of Nondestructive Evaluation}, 42, Article Number 40.

\item
Li, T.-H. (2024) Quantile-crossing spectrum and spline
autoregression estimation. arXiv2412.02513.

\item
Li, T.-H. (2025). Quantile Fourier transform, quantile series, and
nonparametric estimation of quantile spectra. {\it Communications in Statistics - Simulation and
Computation}, DOI: 10.1080/03610918.2025.2509820. 

\item
Lim, Y., and Oh, H.-S. (2022). Quantile spectral analysis of long-memory processes. 
{\it Empirical Economics}, 62, 1245--1266. 

\item
L\"{u}tkepohl, H. (1993). {\it Introduction to Multiple Time Series Analysis},  2nd edn. New York: Springer-Verlag.

\item
Meziani, A., Medkour, T., and Djouani, K. (2020). Penalised quantile periodogram for spectral estimation. 
{\it Journal of Statistical Planning and Inference}, 207, 86--98.

\item
Mokkadem, A. (1988). Mixing properties of ARMA processes. {\it Stochastic Processes and their Applications}, 29,
309--315.

\item
Parzen, E. (1982). Maximum entropy interpretation of autoregressive spectral densities. 
{\it Statistics \& Probability Letters}, 1, 7--11.

\item
Percival, D., and Walden, A. (1993). {\it Spectral Analysis for Physical Applications}, Chap.\ 9.
Cambridge, UK: Cambridge University Press.

\item
R Core Team (2024). R: A language and environment for statistical
  computing. R Foundation for Statistical Computing, Vienna,
  Austria. \url{https://www.R-project.org/}.

\item
Stoica, P., and Moses, R. (1997). 
{\it Introduction to Spectral Analysis}, Chap.\ 3. Upper Saddle River, NJ: Prentice Hall.

\item
Troster, V. (2018). Testing for Granger-causality in quantiles.
{\it Econometric Reviews}, 37, 850--866.

\item
Whittle, P. (1953). Estimation and information in stationary time series. {\it Arkiv f\"{o}r Matematik}, 2, 423--434.

\item
Whittle, P. (1963). On the fitting of multivariate autoregressions, and the approximate canonical
factorization of a spectral density matrix. {\it Biometrika}, 50, 129--134.

\item
Wiener, N., and Masani, P. (1957). The prediction theory of multivariate stochastic processes, I: 
The regularity condition. {\it Acta Mathematica}, 98, 111--150.

\item
Wiener, N., and Masani, P. (1958). The prediction theory of multivariate stochastic processes, II: 
The linear predictor. {\it Acta Mathematica}, 99, 93--137.

\item
Wood, S. (2022). Package `mgcv'. \url{https://cran.r-project.org/web/packages/mgcv/mgcv.pdf}.

\item
Wu, W. (2007). $M$-estimation of linear models with dependent errors. {\it Annals of Statistics}, 35, 495--521.

\end{description}
}

\newpage

\section*{Appendix I: Computation of Spline Autoregression}

\setcounter{equation}{0}
\renewcommand{\theequation}{A.\arabic{equation}}

\setcounter{figure}{0}
\renewcommand{\thefigure}{A.\arabic{figure}}

\setcounter{table}{0}
\renewcommand{\thetable}{A.\arabic{table}}

Recall that any $\bA_\tau(\cdot) \in \cF_m$ can be written as 
$\bA_\tau(\cdot) = \bThe_{\tau} \, \bPhi(\cdot)$. 
Let $\bThe := [\bThe_1,\dots,\bThe_p] \in \bbR^{m \times Kmp}$. Then, with  $\bY_\ell$ and $\bZ_\ell$ defined in Section 4, we have
\eqn
 \sum_{t=p+1}^n
\bigg\| \by_t(\al_\ell) - \sum_{\tau=1}^p  \bA_{\tau}(\al_\ell) \, \by_{t-\tau}(\al_\ell) \bigg\|^2
=  \| \bY_\ell - \bThe \, \bZ_\ell \|^2.
\label{eq1}
\eqqn
Due to the identity $\vect(\bThe \, \bZ_\ell) = (\bZ_\ell^T \otimes \bI_m) \vect(\bThe)$, we can write
\eqn
\| \bY_\ell - \bThe \, \bZ_\ell \|^2 =   \| \by_\ell -\bX_\ell \bmth \|^2
\label{eq1n}
\eqqn
where $\bmth   :=  \vect(\bThe) \in \bbR^{Km^2p}$, $\by_\ell   :=  \vect(\bY_\ell) \in \bbR^{m(n-p)}$, and 
$\bX_\ell  :=  \bZ_\ell^T \otimes \bI_m \in \bbR^{m(n-p)\times Km^2p}$. 
In addition, because $[\bThe_{1} \, \ddot{\bPhi}(\al),\dots,\bThe_{p} \, \ddot{\bPhi}(\al)] 
= \bThe \, ( \bI_p \otimes  \ddot{\bPhi}(\al))$ and $\vect(\bThe \,(\bI_p \otimes  \ddot{\bPhi}(\al)) 
=(  (\bI_p \otimes  \ddot{\bPhi}(\al))^T \otimes \bI_m ) \vect(\bThe)$, we have
\eqn
\sum_{\tau=1}^p \|  \bThe_{\tau} \, \ddot{\bPhi}(\al) \|^2 
 = \| \bThe \, ( \bI_p \otimes  \ddot{\bPhi}(\al)) \|^2
 =  \|  ( \bI_p \otimes  \ddot{\bPhi}^T(\al) \otimes \bI_m ) \bmth \|^2.
\label{eq2}
\eqqn
Substituting (\ref{eq1})--(\ref{eq2}) in (\ref{sar}) leads to the following reformulations
of the SAR problem:
\eqn
\hat{\bThe}   & := & 
\operatorname*{argmin}_{\bThe \in \bbR^{m \times Kmp}} \ \bigg\{
\sum_{\ell=1}^L  \| \bY_\ell -  \bThe \, \bZ_\ell \|^2 
+ (n-p)  \lam \int_{\alu}^{\albar}  \| \bThe \, ( \bI_p \otimes  \ddot{\bPhi}(\al))  \|^2 d\al \bigg\},
\label{sar3a} \\
\hat{\bmth}  & := &
\operatorname*{argmin}_{\bmth \in \bbR^{Km^2p}} \ \bigg\{
\sum_{\ell=1}^L  \| \by_\ell -\bX_\ell \bmth \|^2 
+  (n-p) \lam \int_{\alu}^{\albar}  \| ( \bI_p \otimes  \ddot{\bPhi}^T(\al) \otimes \bI_m )\bmth \|^2 d\al \bigg\}.
\label{sar3b}
\eqqn
The normal equations of (\ref{sar3a}) take the form
\eq
\bThe 
\bigg( \sum_{\ell=1}^L \bZ_\ell \bZ_\ell^T  + (n-p) \lam \bD \bigg)
= \sum_{\ell=1}^L \bY_{\ell} \bZ_{\ell}^T,
\eqq
where $\bD := \bI_p \otimes  \int_{\alu}^{\albar}  \ddot{\bPhi}(\al) \,\ddot{\bPhi}^T(\al) \, d\al$.
The normal equations of (\ref{sar3b}) take the form
\eq
\bigg( \sum_{\ell=1}^L \bX_\ell^T \bX_\ell + (n-p) \lam (\bD \otimes \bI_m) \bigg) \bmth
= \sum_{\ell=1}^L \bX_\ell^T \by_\ell.
\eqq
Therefore, the SAR solution can be expressed as
\eqn
\hat{\bThe}  & = &
\bigg( \sum_{\ell=1}^L \bY_{\ell} \bZ_{\ell}^T \bigg)
\bigg( \sum_{\ell=1}^L \bZ_\ell \bZ_\ell^T  + (n-p) \lam \bD  \bigg)^{\!\!\!-1},
\label{sol3} \\
\hat{\bmth} & = & \bigg( \sum_{\ell=1}^L \bX_\ell^T \bX_\ell + (n-p)\lam (\bD \otimes \bI_m) \bigg)^{\!\!\!-1} 
\bigg(\sum_{\ell=1}^L \bX_\ell^T \by_\ell \bigg).
\label{sol1}
\eqqn
Note that (\ref{sol3}) is more efficient computationally when $m$ is large because it requires 
the inversion of a $Kmp$-by-$Kmp$ matrix, whereas (\ref{sol1}) requires the inversion 
of an $Km^2p$-by-$Km^2p$ matrix. 

The hat (or smoothing) matrix associated with (\ref{sar3b}) is given by
\eq
\bH  :=  \bX_0 \bigg[ \sum_{\ell=1}^L \bX_\ell^T \bX_\ell + (n-p) \lam (\bD \otimes \bI_m) \bigg]^{-1} 
\bX_0^T,
\eqq
where  $\bX_0 := [\bX_1^T,\dots,\bX_L^T]^T$. Therefore,
\eq
\tr(\bH)  & = & \sum_{\ell'=1}^L  \tr \bigg( \bX_{\ell'} \bigg[ \sum_{\ell=1}^L \bX_\ell^T \bX_\ell 
+  (n-p) \lam (\bD \otimes \bI_m) \bigg]^{-1} \bX_{\ell'}^T \bigg) \\
 & = & \sum_{\ell'=1}^L  \tr \bigg( (\bZ_{\ell'}^T \otimes \bI_m)
\bigg[ \bigg( \sum_{\ell=1}^L \bZ_\ell \bZ_\ell^T + (n-p) \lam \bD \bigg)  \otimes \bI_m  \bigg]^{-1}
(\bZ_{\ell'}^T \otimes \bI_m)^T \bigg) \notag \\
& = & 
\sum_{\ell'=1}^L  \tr \bigg( \bZ_{\ell'}^T
\bigg[ \sum_{\ell=1}^L \bZ_\ell \bZ_\ell^T + (n-p) \lam \bD   \bigg]^{-1}
\bZ_{\ell'} \otimes \bI_m \bigg)  \notag \\
& = & \sum_{\ell'=1}^L  \tr \bigg( \bZ_{\ell'}^T 
\bigg[ \sum_{\ell=1}^L \bZ_\ell \bZ_\ell^T + (n-p) \lam \bD   \bigg]^{-1}
\bZ_{\ell'} \bigg) \times m.
\eqq
The GCV criterion in (\ref{gcv}) can be expressed as
\eq
\text{GCV}(\lam) = 
\frac{ (L (n-p))^{-1} \sum_{\ell=1}^L \| \bY_\ell - \hat{\bThe} \bZ_\ell \|^2}
{\{1- (L (n-p))^{-1} \tr(\bH)\}^2}.
\eqq
 By following the convention in {\tt smooth.spline},the smoothing parameter $\lam$ can be 
reparameterized by {\tt spar} such that $\lam =  r \times 256^{3 \times {\tt spar}-1}$, with
$r := (n-p)^{-1}\sum_{\ell=1}^L \tr(\bZ_\ell \bZ_\ell^T)/\tr(\bD)$.

\section*{Appendix II: Granger-Causality Analysis}

We extend the concept of Granger-causality for ordinary AR models
(Granger 1969; L\"{u}kepohl 1993, Section 2.3) to the AR model with functional coefficients 
in (\ref{sar})  for the QSER. We say that $\{ y_{j',t} \}$ is Granger-causal for $\{ y_{j,t} \}$ at quantile level $\al$ if  the $(j,j')$-entry of  $\bA_\tau(\al)$ in (\ref{sar}) is nonzero for some $\tau \in \{1,\dots,p\}$. This causality is related but not identical to the so-called Granger-causality in quantiles  (Chuang et al.\ 2009; Troster 2018; Cheng et al.\ 2022). The latter is defined in terms of quantile regression of the original series. 
The SAR-based Granger-causality is based on least-squares regression of the QSER.

Given the estimates $\{ \hat{\bA}_\tau(\cdot): \tau=1,\dots,p\}$ 
from a data record $\{ \by_t: t=1,\dots,n\}$, one can detect the SAR-based Granger-causality 
by a bootstrap procedure under the assumption that $\{ \by_t(\al) \}$ is 
an AR process with coefficients $\{\bA_\tau(\al): \tau = 1,\dots,p\}$ and  the null hypothesis that
\eq
H_0:  \text{the $(j,j')$-entry of $\bA_\tau(\al)$ equals zero for all $\tau$ and $\al$}. 
\eqq
Specifically, this bootstrap procedure comprises the following steps.

\begin{itemize}
\item[(a)] Let $\{ \bmep_t(\al_\ell): t=1,\dots,n\}$ be the residual series from fitting an  AR$(p)$ 
model to $\{ \by_t(\al_\ell): t=1,\dots,n\}$. Generate 
$\{ \bmep^{(b)}_t (\al_\ell): t=1,\dots,n_B\}$ ($b=1,\dots,B; n_B \gg n$) by sampling the time 
index $\{1,\dots,n\}$ with replacement and rearranging the residuals accordingly.
\item[(b)] Let $\bA_{\tau}^{(0)}(\al_\ell)$ be the same as $\hat{\bA}_\tau(\al_\ell)$ except that 
the $(j,j')$-entry is set to zero for all $\tau$ and $\al_\ell$ under $H_0$. Generate 
$\{ \by_t^{(b)}(\al_\ell): t=1,\dots,n\}$ $(b=1,\dots,B; \ell=1,\dots,L)$ according to
$\by_t^{(b)}(\al_\ell)  := \tilde{\by}_t^{(b)}(\al_\ell) -  n^{-1} \sum_{s=1}^n \tilde{\by}_{s}^{(b)}(\al_\ell)$,
where
\eq
\tilde{\by}_t^{(b)}(\al_\ell) = \sum_{\tau=1}^p 
\bA_\tau(\al_\ell) \tilde{\by}_{t-\tau}^{(b)}(\al_\ell) + \bmep^{(b)}_{n_B-n+t} (\al_\ell) \quad (t=-n_B+n+1,\dots,n)
\eqq
with  $\tilde{\by}_t^{(b)}(\al_\ell) := 0$ for $t=-n_B+n+1-p,\dots,-n_B+n$. The first $n_B-n$ values from this recursion 
are discarded to minimize the effect of initial values. 
\item[(c)] Solve the SAR problem (\ref{sar}) with $\{ \by_t^{(b)}(\al_\ell) \}$ in place of 
$\{ \by_t(\al_\ell)\}$ to obtain $\{ \hat{\bA}_\tau^{(b)}(\cdot): \tau=1,\dots,p\}$ $(b=1,\dots,B)$.
From these samples, construct a pointwise bootstrap confidence band 
for the $(j,j')$-entry of $\hat{\bA}_\tau(\al)$ 
as a function of $\al$ for each $\tau$, and compute the $p$-value of the bootstrap Wald statistic 
(L\"{u}kepohl 1993, Section 3.6)
\eq
W := \hat{\ba}^T \bSig_B^{\dagger} \, \hat{\ba},
\eqq
where $\hat{\ba}$ is the vector formed by the $(j,j')$-entry of $\hat{\bA}_\tau(\al_\ell)$ $(\tau = 1,\dots,p; \ell=1,\dots,L)$ and $\bSig_B^\dagger$ is the generalized inverse of the sample covariance matrix of the vectors 
formed by the $(j,j')$-entry of $\hat{\bA}_\tau^{(b)}(\al_\ell)$ $(\tau = 1,\dots,p; \ell=1,\dots,L)$ for $b=1,\dots,B$.
\end{itemize}
Note that the bootstrap samples in step (a) are serially independent vectors
whose components retain the possible dependence of the residuals for each fixed $t$.
Block sampling schemes may be used as an alternative for handling possible serial dependence.
Unlike Birr et al.\ (2019), the bootstrap samples in our procedure 
are generated for the QSER under the null hypothesis rather than for the original time series.
One may re-center the residuals in step (b) as recommended in Lahiri (2003, p.\ 23), 
but the estimates  in step (c) will not be seriously affected.

\section*{Appendix III: Additional Simulation Results}

\begin{figure}[p]
\centering
\includegraphics[height=5in,angle=-90]{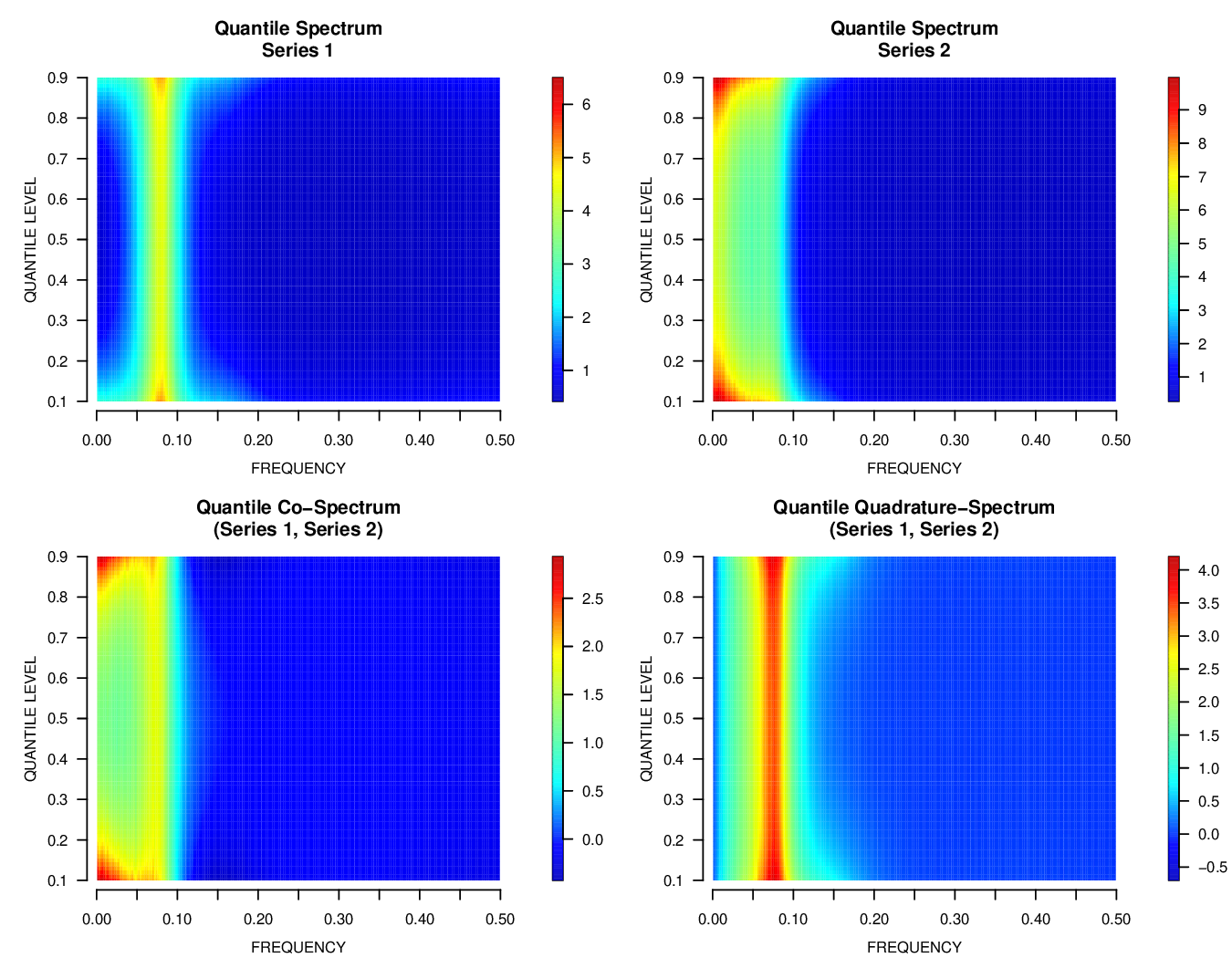}  
\caption{Quantile spectrum of the ARMA process  (\ref{arma21}). }
 \label{fig:qspec:arma}
\vspace{0.2in}
\centering
\includegraphics[height=5in,angle=-90]{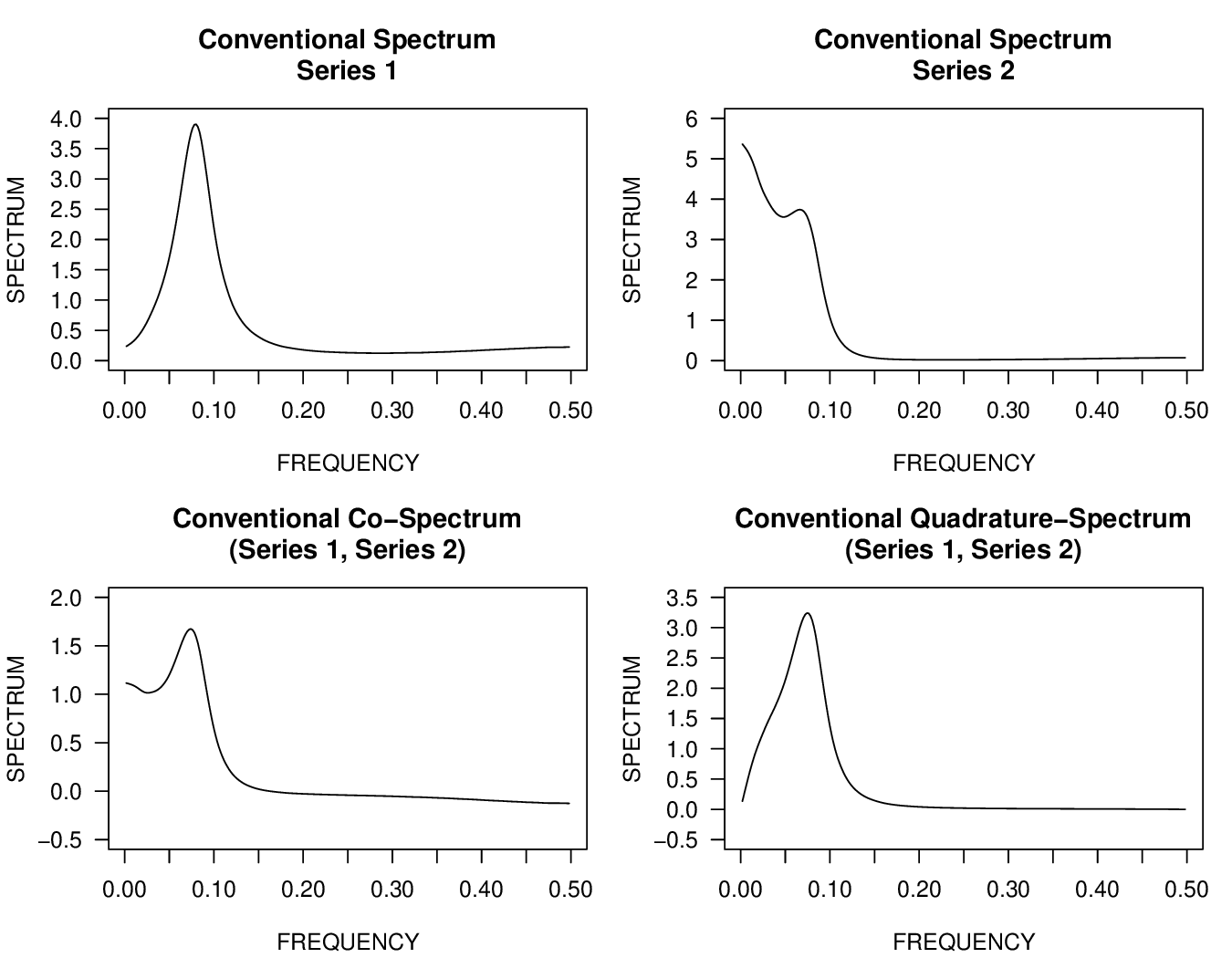}  
\caption{Conventional spectrum of the ARMA process  (\ref{arma21}). }
 \label{fig:spec:arma}
\end{figure}

Consider  the ARMA process $\by_t := [y_{1,t},y_{2,t}]^T$ defined by
\eqn
\by_t - \bA_1 \, \by_{t-1} - \bA_2 \, \by_{t-2} =  \bmep_t + \bB \, \bmep_{t-1},  \quad \{\bmep_t \} \sim \text{ IID } \N(\0,\bSig),
\label{arma21}
\eqqn
where 
\eq
\bA_1 & := & \left[
\begin{array}{rr}
0.816 & 1.246 \\
0.558 & 1.107
\end{array} 
\right], \quad 
\bA_2 := \left[
\begin{array}{rr}
0.643 & 1.184 \\
0.307 & 0.203
\end{array} 
\right], \\
\bB & := & \left[
\begin{array}{rr}
0 & 2.496 \\
0.4 & 0
\end{array} 
\right], \quad
 \bSig := \left[
\begin{array}{rr}
0.04 & -0.02 \\
-0.02 & 0.02
\end{array} 
\right].
\eqq
Figure~\ref{fig:qspec:arma} depicts the quantile spectrum of this process.
Figure~\ref{fig:spec:arma} depicts the conventional spectrum of this process.

\begin{table}[H]
{\footnotesize
\begin{center} 
\caption{Mean KLD   of Spectral Estimators for the ARMA Process (\ref{arma21})}
\label{tab:err:arma2}
\begin{tabular}{cccccccccccc} \toprule
 &  \multicolumn{2}{c}{SAR} &  \multicolumn{3}{c}{AR} &  \multicolumn{3}{c}{LW} \\
 $n$ &  GCV    &  Fixed {\tt spar}  &  None & SPLINE  & GAMM   & None & SPLINE  & GAMM \\   \midrule
256 &   0.150 & 0.110 & 0.250 & 0.246  &  0.149 & 0.289 & 0.263 & 0.171 \\
512 &  0.083 & 0.068   & 0.170 & 0.167 & 0.106 & 0.182  & 0.179 & 0.106 \\
\hline
\end{tabular} 
\end{center}
}
{\scriptsize 
\begin{center} 
\begin{minipage}{5in}
Results are based on 1000 Monte Carlo runs. 
Fixed {\tt spar}:  {\tt spar}  = 1 for $n=256$ and {\tt spar} = 0.9 for $n=512$.
SPLINE:  {\tt smooth.spline}. GAMM: {\tt gamm} with correlated residuals. LW: lag-window estimator using Tukey-Hanning window with optimal bandwidth parameter $M=14$ for $n=256$ and $M=17$ for $n=512$.
\end{minipage}
\end{center}
}
\begin{figure}[H]
\centering
\vspace{0.3in}
\includegraphics[height=3in,angle=-90]{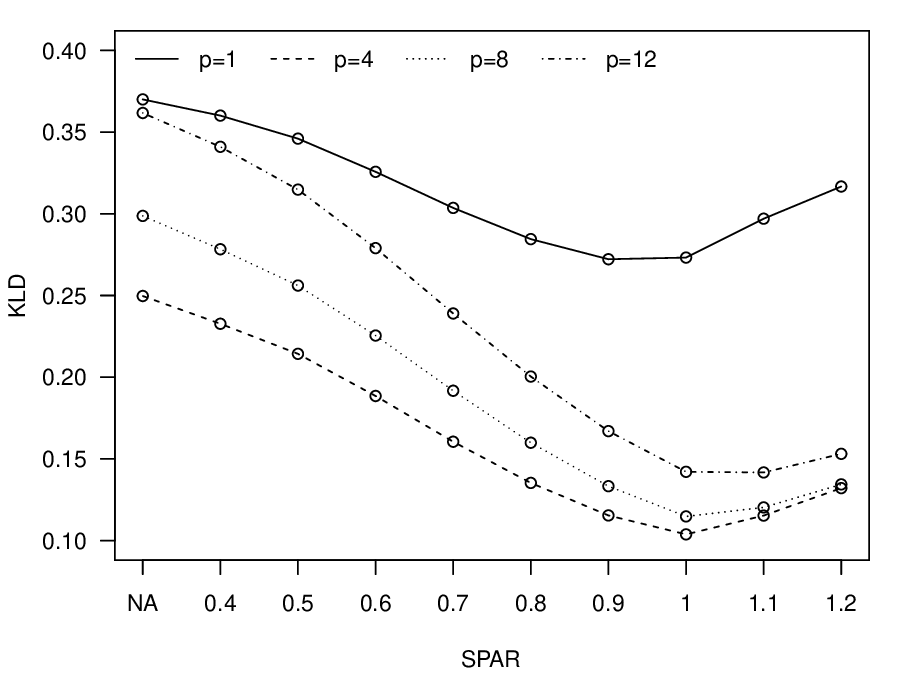}
\includegraphics[height=3in,angle=-90]{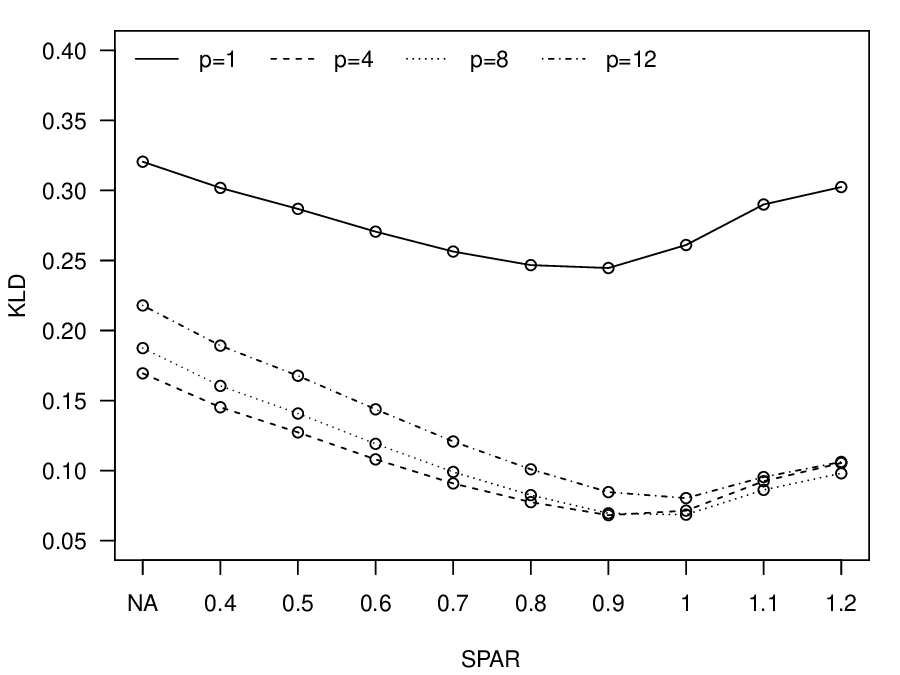}
\centerline{\hspace{0.3in}(a) \hspace{2.8in}(b)}
\caption{Mean KLD of the SAR estimator for the ARMA process (\ref{arma21}) with different choices of $p$ and {\tt spar} ({\tt spar} = NA corresponds to the AR estimator without quantile smoothing). (a) $n=256$. (b) $n=512$.
Results are based on 1000 Monte Carlo runs}
\label{fig:err:arma2}
\end{figure}
\end{table}

The results of spectral estimation  for this process are shown in Table~\ref{tab:err:arma2} and Figure~\ref{fig:err:arma2}.  As can be seen in Table~\ref{tab:err:arma2}, the SAR estimator outperforms
 the AR and LW estimators with or without quantile smoothing. The mean KLD of the SAR 
estimator with the smoothing parameter selected by GCV is reasonably close 
to the best values achieved with fixed  {\tt spar} (Figure~\ref{fig:err:arma2}).
These findings are similar to the findings in Section 4 for the mixture process (\ref{y}).

\begin{figure}[H]
\centering
\vspace{0.2in}
\includegraphics[height=5in,angle=-90]{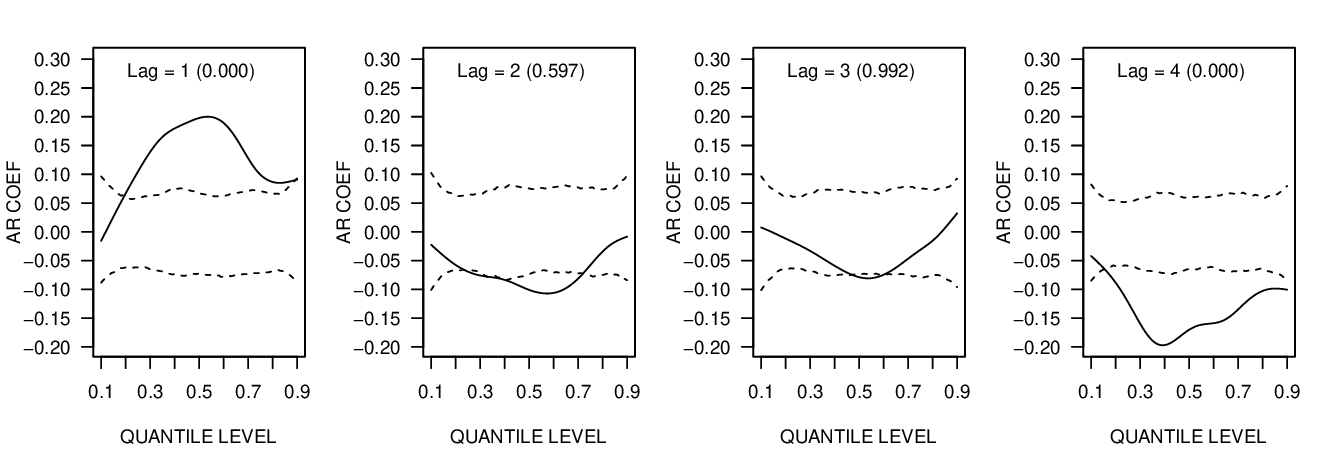}  
\caption{The (1,2)-entry of $\hat{\bA}_\tau(\cdot)$ at $\tau=1,\dots,4$
for  a series generated from the ARMA process in (\ref{arma21}) with $n=512$.
Dashed lines depict the pointwise 95\% bootstrap confidence band constructed from 
1000 bootstrap samples under the hypothesis of no causality. Numbers in parentheses are $p$-values of 
the bootstrap Wald statistic.}
\label{fig:causality:arma}
\end{figure}

The results of SAR-based Granger-causality analysis for the ARMA process in (\ref{arma21}) are shown in Figure~\ref{fig:causality:arma} and Table~\ref{tab:causality:arma}. 
These results confirm the existence of Granger-causality of series 2 for series 1 
as expected from (\ref{arma21}). The effect of this causality is largely confined 
to $\tau=1$ and $\tau=4$. It is stronger in the middle quantile region than in the tail regions.

\begin{table}[H]
{\footnotesize
\begin{center} 
\caption{Mean $p$-Values of Wald Test on (1,2)-Entry  
for the ARMA Process (\ref{arma21})}
\label{tab:causality:arma}
\begin{tabular}{lccccccccccccc} \toprule
 $\tau=1$ & $\tau=2$ & $\tau=3$ & $\tau=4$ & All $\tau$ \\  \midrule
 {\bf 0.001} & 0.601& 0.453& 0.128& {\bf 0.000} \\
\hline
\end{tabular} 
\end{center}
}
{\scriptsize 
\begin{center}
\begin{minipage}{2.3in}
Results are based on 1000 Monte Carlo run $(n=512)$. The $p$-value in each run is computed from 1000 bootstrap samples. Boldface font highlights the cases where $p$-value is less than 0.05.
\end{minipage}
\end{center}
}
\end{table}

\newpage
\section*{Appendix IV: R Functions}

The following functions in the R package `qfa' (version $\ge$ 3.1) are implementations of the SAR method for quantile spectral estimation and Granger-causality analysis. The package is available for installation at 
\url{cran.r-project.org} and \url{github.com/thl2019/QFA}.

\begin{itemize}
\item {\tt qdft}: a function that computes the QDFT at a given sequence of quantile levels from a (univariate or multivariate) time series.
\item {\tt qper}: a function that computes the quantile periodogram at a given sequence of quantile levels 
from a time series or the QDFT.
\item {\tt qser}: a function that computes the QSER at a given sequence of quantile levels from 
a time series or the QDFT.
\item {\tt qacf}: a function that computes the QACF at a given sequence of quantile levels from 
a time series or the QDFT.
\item {\tt qspec.ar}: a function that computes the AR spectral estimate at a given sequence of quantile levels with or without subsequent quantile smoothing from a time series or the QSER.
\item {\tt qspec.sar}: a function that fits the SAR spectral estimate at a given sequence of quantile levels with or without subsequent quantile smoothing from a time series or the QSER.
\item {\tt qspec.lw}: a function that computes the LW spectral estimate at a given sequence of quantile levels with or without  quantile smoothing from a time series or the QACF.
\item {\tt qfa.plot}: a function that produces an image plot for a real-valued spectrum as a function of frequency and quatile level
\item {\tt qkl.divergence}: a function that computes the Kullback-Leibler divergence (KLD) of a spectral estimate against the true spectrum.
\item {\tt sar.gc.coef}: a function that extracts the functional AR coefficients of the SAR model produced by {\tt qspec.sar}.
\item {\tt sar.gc.bootstrap}: a function that generates bootstrap samples of functional AR coefficients from the SAR model produced by {\tt qspec.sar} for Granger-causality analysis.
\item {\tt sar.gc.test}: a function that computes the bootstrap Wald statistic and its $p$-value for Granger-causality,  together with 
the 95\% confidence band,  from the functional AR coefficients produced by {\tt sar.gc.coef} and the corresponding bootstrap samples produced by {\tt sar.gc.bootstrap}.
\end{itemize}

\end{document}